\documentclass{aa}
\usepackage[varg]{txfonts}
\usepackage{graphicx}
\usepackage{natbib}
\usepackage{subfigure}
\usepackage{dcolumn}
\usepackage{lscape}
\bibpunct{(}{)}{;}{a}{}{,}

\begin{document}

\title{The Fornax Deep Survey (FDS) with VST \\
VI. Optical properties of the dwarf galaxies in the Fornax cluster }

\author{Aku Venhola\inst{1}$^,$ \inst{2} 
	\and Reynier Peletier \inst{2} 
	\and Eija Laurikainen\inst{1} 
	\and Heikki Salo\inst{1} 
	\and Enrichetta Iodice\inst{3}
	\and Steffen Mieske \inst{4}
	\and Michael Hilker \inst{5}
	\and Carolin Wittmann \inst{6}
	\and Maurizio Paolillo \inst{7,3}
	\and Michele Cantiello \inst{8}
	\and Joachim Janz \inst{1,13}
	\and Marilena Spavone \inst{3}
	\and Raffaele D'Abrusco \inst{9}
	\and Glenn van de Ven \inst{14}
	\and Nicola Napolitano \inst{3}
	\and Gijs Verdoes Kleijn \inst{2}
	\and Massimo Capaccioli \inst{10}
	\and Aniello Grado \inst{3}	
	\and Edwin Valentijn \inst{2}
	\and Jes\'us Falc\'on-Barroso \inst{11,12}
	\and Luca Limatola \inst{3}
}

\offprints{A. Venhola, \email{aku.venhola@oulu.fi}}

\institute{Astronomy Research Unit, University of Oulu, Pentti Kaiteran katu 1, 90014 Oulu, Finland 
  \and Kapteyn Institute, University of Groningen, Landleven 12, 9747 AD Groningen, the Netherlands
  \and INAF - Astronomical Observatory of Capodimonte, Salita Moiariello 16, I80131, Naples, Italy 
  \and European Southern Observatory, Alonso de Cordova 3107, Vitacura,  Santiago, Chile
  \and European Southern Observatory, Karl-Schwarzschild-Strasse 2, D-85748 Garching bei München, Germany 
  \and Astronomisches Rechen-Institut, Zentrum f\"ur Astronomie der Universit\"at Heidelberg, M\"onchhofstra\ss e 12-14, 69120 Heidelberg, Germany
  \and University of Naples Federico II, C.U. Monte Sant'Angelo, Via Cinthia, 80126 Naples, Italy
  \and INAF Osservatorio Astronomico d'Abruzzo,  Via Maggini, 64100, Teramo, Italy
  \and Smithsonian Astrophysical Observatory, 60 Garden Street, 02138 Cambridge, MA, USA
  \and  University of Naples Federico II, C.U. Monte Sant'Angelo, Via Cinthia, 80126 Naples, Italy
  \and Instituto de Astrofisica de Canarias, C/ Via L'actea s/n, 38200 La Laguna, Spain 
  \and Depto. Astrofisica, Universidad de La Laguna,  C/ Via L'actea s/n, 38200 La Laguna, Spain
  \and Finnish Centre of Astronomy with ESO (FINCA)
, University of Turku, Väisäläntie 20, FI-21500 Piikkiö, Finland 
  \and Department of Astrophysics, University of Vienna, T\"urkenschanzstrasse 17, 1180 Wien, Austria }

\date{Received \today / Accepted April 10,2019}

\abstract { Dwarf galaxies are the most common type of galaxies in galaxy clusters. Due to their low mass, they are more vulnerable to environmental effects than massive galaxies, and are thus optimal for studying the effects of the environment on galaxy evolution. By comparing the properties of dwarf galaxies with different masses, morphological types, and cluster-centric distances we can obtain information about the physical processes in clusters that play a role in the evolution of these objects and shape their properties. The Fornax Deep Survey Dwarf galaxy Catalog (FDSDC) includes 564 dwarf galaxies in the Fornax cluster and the in-falling Fornax A subgroup. This sample allows us to perform a robust statistical analysis of the structural and stellar population differences in the range of galactic environments within the Fornax cluster.
} 
{ By comparing our results with works concerning other clusters and the theoretical knowledge of the environmental processes taking place in galaxy clusters, we aim to understand the main mechanisms transforming galaxies in the Fornax cluster.
 }
{
 We have exploited the FDSDC to study how the number density of galaxies, galaxy colors and structure change as a function of the cluster-centric distance, used as a proxy for the galactic environment and in-fall time. We also used deprojection methods to transform the observed shape and density distributions of the galaxies into the intrinsic physical values.  These measurements are then compared with predictions of simple theoretical models of the effects of harassment and ram pressure stripping on galaxy structure. We used stellar population models to estimate the stellar masses, metallicities and ages of the dwarf galaxies. We compared the properties of the dwarf galaxies in Fornax with those in the other galaxy clusters with different masses.
}
{We present the standard scaling relations for dwarf galaxies, which are the size-luminosity, S\'ersic $n$-magnitude and color-magnitude relations. New in this paper is that we find a different behavior for the bright dwarfs (-18.5 mag < M$_{r'}$ < -16 mag) as compared to the fainter ones (M$_{r'}$ > -16 mag): While considering galaxies in the same magnitude-bins, we find that, while for fainter dwarfs the g'-r' color is redder for lower surface brightness objects (as expected from fading stellar populations), for brighter dwarfs the color is redder for the higher surface brightness and higher S\'ersic $n$ objects. The trend of the bright dwarfs might be explained by those galaxies being affected by harassment and by slower quenching of star formation in their inner parts. As the fraction of early-type dwarfs with respect to late-types increases toward the central parts of the cluster, the color-surface brightness trends are also manifested in the cluster-centric trends, confirming that it is indeed the environment that changes the galaxies. We also estimate the strength of the ram-pressure stripping, tidal disruption, and harassment in the Fornax cluster, and find that our observations are consistent with the theoretically expected ranges of galaxy properties where each of those mechanisms dominate. We furthermore find that the luminosity function, color-magnitude relation, and axis-ratio distribution of the dwarfs in the center of the Fornax cluster are similar to those in the center of the Virgo cluster. This indicates that in spite of the fact that the Virgo is six times more massive, their central dwarf galaxy populations appear similar in the relations studied by us.} {}


\maketitle

\section{Introduction}

%\doublespacing
The local density of galaxies has been shown to play an important role in galaxy evolution, leading the more quiescent galaxies to appear preferentially in high density regions in the local Universe (\citealp{Dressler1980}, \citealp{PengY2010}). This is the case for high mass galaxies and the tendency is even stronger for low mass galaxies \citep{Binggeli1990}. A range of physical processes has been suggested to be responsible for such environmental variance (\citealp{Boselli2008}, \citealp{Serra2012}, \citealp{Jaffe2018}, \citealp{Moore1998},\citealp{Mastropietro2005},\citealp{Rys2014}, \citealp{Toloba2015}). However, the properties of galaxies do not only depend on their surroundings, but also on their mass. For example, massive galaxies have been shown to host on average older stellar populations (see e.g., \citealp{Thomas2005}, \citealp{Roediger2017}, \citealp{Schombert2018}), and to have more substructure than the less massive galaxies (\citealp{Herrera-Endoqui2015}, \citealp{Janz2014}). In order to isolate the environmental effects from internal processes, we need to study the galaxies over a range of mass bins and environments. For such a study, dwarf galaxies  constitute an important resource. Dwarf elliptical galaxies (dE) are the most abundant galaxies in galaxy clusters. They have low masses and low surface brightnesses, making them relatively easily affected by the environment. Thus, due to their abundance and vulnerability, they can be used to study the effects of environment on galaxy evolution.

\indent Studies concentrating on mid- and high-redshift galaxies have shown an increase in the fraction of massive red quiescent galaxies in clusters since z = 1-2 up to present (\citealp{Bell2004}, \citealp{Cassata2008}, \citealp{Mei2009}). The emergence of these quiescent galaxies is partly explained by their mass: massive galaxies form stars more efficiently (\citealp{Pearson2018}) and therefore, if no more gas is accreted, run out of their cold gas reservoir faster than less massive galaxies. Additionally, the internal energy of gas and stars in these galaxies increases due to merging of satellite galaxies (\citealp{DiMatteo2005}), and due to feedback from active galactic nuclei (AGN) and supernovae (\citealp{Bower2006}, \citealp{Hopkins2014}), which prevents the gas from cooling and collapsing into new stars. The ensemble of these processes is called mass quenching since they are all related to the mass of the galaxy's dark matter halo, rather than to the environment. If the formation of early-type galaxies would happen only via mass quenching, we should see only very massive early-type galaxies at high redshift, and an increasing number of lower mass early-type galaxies toward lower redshift (\citealp{Thomas2005}). Indeed, observations have confirmed such evolution among the quiescent galaxies (\citealp{Bundy2006}) showing that mass-quenching is an important mechanism in galaxy evolution. However, strong correlations are also found with regard to stellar populations and galaxy morphology (\citealp{PengY2010}) with the environment, indicating that mass quenching is not the only process contributing to the formation of early-type systems. Additionally, studies analyzing isolated early-type galaxies (\citealp{Geha2012}, \citealp{Janz2017}), for which the effects of the external processes can be excluded, have shown that all such quiescent galaxies are more massive than 1$\times$10$^9$ M$_{\odot}$, indicating that mass-quenching is only effective in the mass range larger than that.

\indent In dense environments, such as galaxy clusters, galaxies experience environmental processes acting both on their stellar and gas components. Harassment (\citealp{Moore1998}) is a term used for the high velocity tidal interactions between cluster galaxies, and between a galaxy and the cluster potential, that tend to rip out stellar, gas and dark matter components of the galaxies. \citet{Mastropietro2005} showed that harassment can transform a disk galaxy into a dynamically hot spheroidal system in time scales of a few Gyrs. Additionally, these processes may trigger bursts of star formation that quickly consume the cold gas reservoir, transforming the galaxy into an early-type system (\citealp{Fujita1998}).  Another consequence of harassment is that galaxies should become more compact as their outer parts are removed \citep{Moore1998}. Indeed, \citet{Janz2016} found that the sizes of field late-type dwarfs are twice as large as those of early and late-type galaxies of the same mass located in the Virgo cluster. However, \citet{Smith2015} showed that harassment is effective in truncating the outer parts of only galaxies that have highly radial orbits with short peri-center distances with respect to the cluster center. This challenges the idea that harassment alone can explain the difference between the galaxies in different environments. 

\indent Another, more drastic consequence of the gravitational interactions is that in the cluster center they are so strong that they may cause complete disruption of dwarf galaxies (\citealp{McGlynn1990}, \citealp{Koch2012}). The material of those disrupted galaxies then ends up in the intra-cluster medium, piling up in the central regions of the cluster, and leaving an underdense core in the galaxy number density profile of the cluster. In the Fornax cluster, which is the object of this study, such a core has been observed (\citealp{Ferguson1989}). Also ultracompact dwarf galaxies, of which some are likely to be remnant nuclei of stripped dwarf galaxies are detected (\citealp{Drinkwater2003}, \citealp{Voggel2016}, \citealp{Wittmann2016}), which gives further evidence for disruption of the dwarfs. Recently, \citet{Iodice2018} studied massive early-type galaxies (ETGs) in the central region of the Fornax cluster, and showed that in the core of the cluster, the ETGs tend to have twisted and/or asymmetric outskirts, and that patches of intra-cluster light ({\citealp{Iodice2016}, \citealp{Iodice2017b}) have been also been detected in that region. These findings suggest that in the core of the Fornax cluster, harassment is shaping the massive galaxies. Also \citet{DAbrusco2016} have identified a population of intra-cluster globular clusters (GCs) that may originate from globular clusters stripped out of the cluster galaxies.

\indent Another process affecting galaxies in clusters, but acting only on their gas component, is ram-pressure stripping \citep{GunnGott1972}. If a gas-rich galaxy moves through a galaxy cluster that contains a significant amount of hot gas, pressure between the gas components can remove the cold gas from the galaxy's potential well. This happens especially on galaxy orbits near the center of the cluster, in which case all the gas of the galaxy might be depleted in a single crossing, corresponding to a time scale of half a Gyr \citep{Lotz2018}. If not all the gas of the galaxy is removed at once, it is likely that the remaining gas is located in its center where the galaxy's anchoring force is strong. This may lead to younger stellar populations in the central part of the dwarf galaxies. Indeed, many dwarf galaxies in clusters have blue centers (\citealp{Lisker2006}), indicative of their recent star formation. There is strong observational evidence about the removal of gas from galaxies when they enter the cluster environment: Long H$_{\mathrm{ {\sc I}}}$ tails have been observed trailing behind late-type galaxies falling into clusters (\citealp{Kenney2004}, \citealp{Chung2009}). Using H$\alpha$- and UV-imaging, ram pressure stripping has been observed in action in infalling galaxies (\citealp{Poggianti2017}, \citealp{Jaffe2018}). Also, galaxies in clusters have shown to be mostly gas-poor, and those outside the clusters gas-rich (\citealp{Solanes2001}, \citealp{Serra2012}). 

\indent As discussed above, both harassment and ram-pressure stripping take place in the cluster environment. The physical theory behind them is well understood. However, what is not known, is the importance of these processes in shaping the dwarf galaxy populations in clusters. Our understanding of this problem is limited on the theoretical side by the resolution of cosmological simulations, which are not able to resolve small galaxies in the cluster environment (\citealp{Pillepich2018}, \citealp{Schaye2015}). Observationally, it is due to the lack of systematic studies analysing the properties of dwarf galaxies for a complete galaxy sample, spanning a range of different galactic environments. Recently, some surveys including the Next Generation Fornax Survey (NGFS, \citealp{Eigenthaler2018}) and the Fornax Deep  Survey (FDS, \citealp{Iodice2016}) have been set up to address these questions. In this study, we approach this problem from the observational point of view using a size and luminosity limited sample of galaxies in the Fornax cluster. This sample is used to test some first order predictions arising from the current theoretical understanding of the different environmental processes.

\indent The Fornax Deep Survey (FDS) provides a dataset that allows to study the faintest dwarf galaxies over the whole Fornax cluster area. The survey covers a 26 deg$^2$ area of the cluster up to its virial radius, covering also the non-virialized southwest Fornax A sub-group. The deep u', g', r', i'-data with wide dynamical range allows us to study the galaxy properties and colors down to the absolute r'-band magnitude M$_{r'}$ $\approx$ -10 mag. The Fornax Deep Survey Dwarf galaxy Catalog (FDSDC) by Venhola et. al. (2018) includes 564 dwarf galaxies in the survey area, and reaches 50\% completeness at the limiting magnitude of M$_{r'}$ = -10.5 mag, and the limiting mean effective surface brightness of $\bar{\mu}_{e,r'}$ = 26 mag arcsec$^{-2}$. The minimum semi-major axis size\footnote{Semi-major axes from SExtractor \citep{Bertin1996} were used.} is $a$ = 2 arcsec. Our sample is obtained from data with similar quality to that of the Next Generation Virgo Survey (NGVS, \citealp{Ferrarese2012}), thus allowing a fair comparison between the faint galaxy populations in these two environments. 

\indent In this paper, our aim is to use the FDSDC to study how the galaxy colors, the luminosity function, the fraction of morphological types, and the structure parameters change from the outer parts of the cluster toward the inner parts. If harassment is the main environmental effect acting on galaxies, the galaxies are expected to become more compact (higher S\'ersic indices and smaller effective radii) and rounder in the inner parts of the cluster, as their outer parts get stripped and their internal velocity dispersion increases (\citealp{Moore1998}, \citealp{Mastropietro2005}). On the other hand, if the main mechanism is ram pressure stripping, the galaxies should not become more compact or rounder toward the cluster center. However, they are expected to become redder and to have lower surface brightnesses after the ram pressure stripping has quenched their star formation. The optical region where the contribution of young stars is significant, is particularly favorable for this kind of studies. We also compare our results with those obtained in the Virgo cluster by NGVS.

\indent In this paper, we first discuss the data and the sample in Section 2. In Sections 3, 4, 5 and 6, we present the distribution, luminosity function, structure, and colors of the dwarf galaxies in the Fornax cluster as a function of their cluster-centric distance and total luminosity, respectively. Finally, in Sections 7 and 8 we discuss the results and summarize the conclusions, respectively. Throughout the paper we use the distance of 20.0 Mpc for the Fornax cluster corresponding to the distance modulus of 31.51 mag \citep{Blakeslee2009}. At the distance of the Fornax cluster, 1 deg corresponds to 0.349 Mpc.

\section{Data}

\subsection{Observations}

We use the FDS, which consists of the combined data of Guaranteed Time Observation Surveys, FOrnax Cluster Ultra-deep Survey (FOCUS, P.I. R. Peletier) and VST Early-type GAlaxy Survey (VEGAS, P.I. E. Iodice), dedicated to the Fornax cluster. Both surveys are performed with the ESO VLT Survey Telescope (VST), which is a 2.6-meter diameter optical survey telescope located at Cerro Paranal, Chile \citep{Schipani2012}. The imaging is done in the u', g', r' and i'-bands using the $1^{\circ} \, \times \, 1^{\circ}$ field of view OmegaCAM instrument \citep{Kuijken2002} attached to VST. The observations cover a 26 deg$^2$ area centered to the Fornax main cluster and the in-falling Fornax A sub-group in which NGC 1316 is the central galaxy (see Fig. \ref{fig:galaxy_locations}). Full explanations of the observations and the AstroWISE-pipeline \citep{McFarland2013} reduction steps are given in Peletier et al. ({\it in prep.}), and \citet{Venhola2018}.

\begin{figure*}
	\includegraphics[width=17cm]{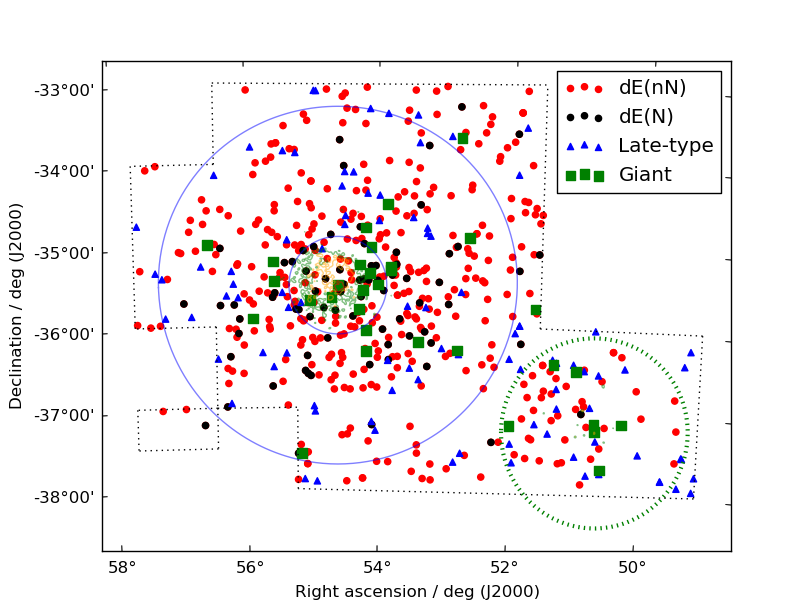}
    \caption{Locations of the likely cluster galaxies in the FDSDC (Venhola et al., 2018). The red, black and blue symbols correspond to non-nucleated dwarf ellipticals, dE(nN), nucleated dwarf ellipticals, dE(N) and late-type dwarfs, respectively. The black dotted lines show the extent of the FDS observations. The inner and outer blue circles show the core (\citealp{Ferguson1989}) and virial radius (\citealp{Drinkwater2001}) of the Fornax main cluster respectively, and the green dotted circle encloses an area within one degree from the Fornax A sub-group (\citealp{Drinkwater2001}). The green, yellow, and red contours around the center of the cluster and Fornax A correspond to the 0.96, 1.15, and 1.34 $\times$ 10$^{-3}$ counts arcmin$^{-2}$ sec$^{-1}$ isophotes of the ROSAT X-ray observations, respectively (\citealp{Kim1998}, \citealp{Paolillo2002}). In the figure north is up and east is toward left.}
   	\label{fig:galaxy_locations}
\end{figure*}

\indent The photometric errors related to the zeropoint calibration of the data were defined by Venhola et al. (2018), and they are 0.04, 0.03, 0.03, and 0.04 mag in u', g', r' and i', respectively. The completed fields are covered with homogeneous depth with the 1$\sigma$ limiting surface brightness over one pixel area of 26.6, 26.7, 26.1 and 25.5 mag arcsec$^{-2}$ in u', g', r' and i', respectively. When averaged over one arcsec$^2$ area, these numbers correspond to 28.3, 28.4, 27.8, 27.2 mag arcsec$^{-2}$ in u', g', r' and i', respectively. The data is calibrated using AB-magnitude system.

\indent The FDS has provided two galaxy catalogs: The catalog containing all dwarf galaxies in the whole FDS area has been presented in \citet{Venhola2018}, and the photometry of the ETGs in the central 9 deg$^2$ area of the cluster by \citet{Iodice2018}. This study mostly considers only the dwarf galaxies, but in Section 9,we also use the stellar masses of the ETGs from the catalog of Iodice et al.

\subsection{Galaxy sample}

We use the FDSDC (Venhola et al., 2018) that is based on the whole data set of the FDS. In Venhola et al. 2018, we ran SExtractor \citep{Bertin1996} on the stacked g'+r'+i'-images to detect the objects, and then selected those with semi-major axis lengths larger than  $a$ > 2 arcsec for further analysis. The galaxy catalog contains the Fornax cluster dwarfs with M$_{r'}$ > $-18.5$ mag, and reaches its 50\% completeness limit at the r'-band magnitude M$_{r'}$ = $-10.5$ mag and the mean effective surface brightness $\bar{\mu}_{e,r'}$ = 26 mag arcsec$^{-2}$.  In Appendix A, we show how the minimum size and magnitude limits affect the completeness of the FDSDC in the size-magnitude space.

\indent In Venhola et al. (2018), we produced S\'ersic model fits for these galaxies using GALFIT \citep{Peng2002,Peng2010} and the r'-band images of the FDS. For the dwarf galaxies with nuclear star clusters we fit the unresolved central component with an additional PSF component. We also measured aperture colors within the effective radii using elliptical apertures. The effective radius, position angle and ellipticity of the apertures were taken from the r'-band S\'ersic fits. For the total magnitudes and the other structural parameters, we use the values adopted from the same r'-band fits.  In the u'-band color analysis, we exclude galaxies that have $\bar{\mu}_{e,r'}$ > 25 mag arcsec$^{-2}$, since their colors are uncertain due to the low signal-to-noise.

\indent We analyze all the galaxies that we classified as likely cluster members in \citet{Venhola2018}. We also use the same division of galaxies into late- and early-types as was used in that work. This morphological division was made visually by classifying the red and smooth galaxies as early-types, and the galaxies that have star formation clumps as late-types. For the low mass galaxies for which the visual morphological classification was uncertain, we used a combination of colors, concentration parameter $C$, and Residual Flux Fraction $RFF$, to separate them into the different types. $RFF$ \citep{Hoyos2011} measures the magnitude of the absolute residuals between the galaxy's light profile and a best-fitting S\'ersic profile, and $C$ \citep{Conselice2014} is defined using the ratio of the galactocentric radii that enclose 80\% and 20\% of the galaxy's light. Details of the morphological classifications and measurements are explained in \citet{Venhola2018}. Due to a lack of more detailed morphological classifications of the FDSDC galaxies, in this study we refer to the late-type dwarfs as dwarf irregulars, dIrrs, and early-type dwarfs as dwarf ellipticals, dEs. We emphasize that this leads into those classes consisting of several types of objects, for example that following the classification system of \citet{Kormendy2012}, spheroidals and "true dwarf ellipticals" would all be classified as dEs in this work, and the irregulars would belong to dIrr. 

\indent Our total galaxy sample in this work consists of 564 Fornax cluster dwarf galaxies. Of these 470 are classified as early-types and 94 as late-types. The early-types are further divided into nucleated, dE(N) (N=81), and non-nucleated, dE(nN) (N=389), dwarf ellipticals. Our sample is at least 95 \% complete within the given limits. Since most of the sample  galaxies have no spectroscopic confirmations, it is likely that there is some amount of galaxies that are actually background objects. According to the estimation of \citet{Venhola2018}, there might be at maximum 10\% background contamination in the sample, corresponding to few tens of misclassified background galaxies.

\subsection{Stellar masses}

We estimate the stellar mass of the sample galaxies using the empirical relation observed between the g'-i' color and mass-to-light ($M/L$) ratio by \citet{Taylor2011}:

\begin{equation}
\log_{10} \left( \frac{ \mathrm{M}_{*}}{\mathrm{M}_{\odot}}\right) = 1.15 + 0.70\times(g'-i') - 0.4\mathrm{M}_{i'}.
\end{equation}

\noindent In the equation, M$_{i'}$  is the absolute i'-band magnitude. As we have done the Se\'rsic model fitting only for the r'-band images, and thus do not have the i'-band total magnitudes for the galaxies, we transform the r'-band total magnitudes into i'-band, by applying the r'-i' color (measured within R$_e$) correction:

\begin{equation}
\log_{10} \left( \frac{ \mathrm{M}_{*}}{\mathrm{M}_{\odot}}\right) = 1.15 + 0.70\times(g'-i') - 0.4\mathrm{M}_{r'} + 0.4\times(r'-i').
\end{equation}

The relation between the stellar masses and r'-band magnitudes for the galaxies in our sample are shown with the black symbols in Fig. \ref{fig:stellar_mass}. According to \citet{Taylor2011} the accuracy of the transformed stellar masses is within 0.1 dex. However, their sample does not contain galaxies with M$_*$ < 10$^{7.5}$ M$_{\odot}$, where we have to use an extrapolation of their relation to obtain the lowest stellar masses. This caveat makes our mass estimations for the lowest mass galaxies uncertain. For quantifying that uncertainty, we estimate the masses also using another method. 

\begin{figure}
	\resizebox{\hsize}{!}{\includegraphics{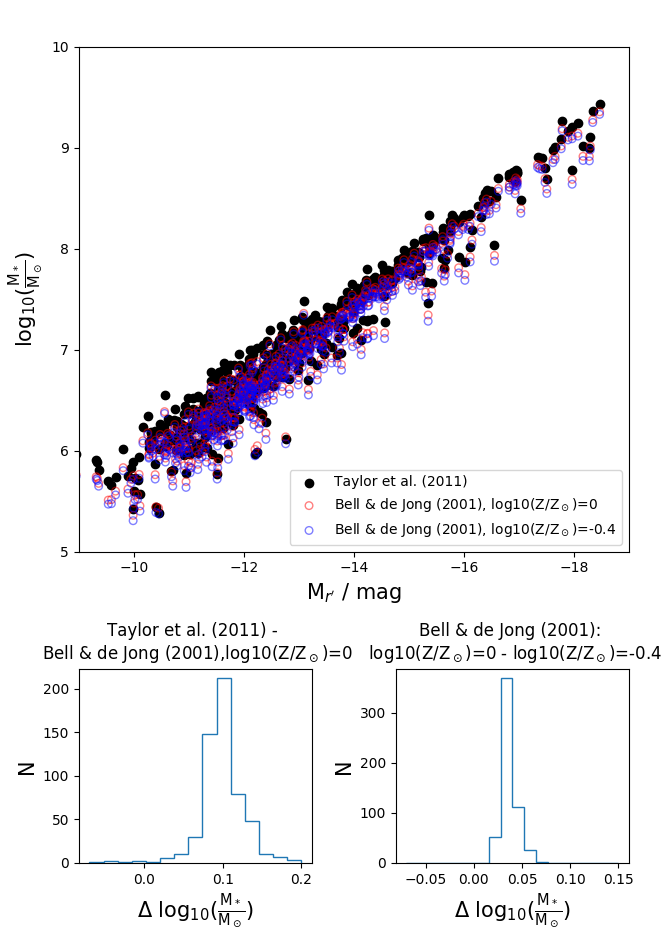}}
    \caption{Upper panel shows stellar masses of the galaxies as a function of their r'-band absolute magnitude (M$_{r'}$). The black symbols show the transformations when using the relations of  \citet{Taylor2011} and the red and blue symbols show the relations when using Bell and de Jong (2001) transformations with the assumed metallicities of 0 and -0.4, respectively. The lower left panel shows a histogram of the differences between stellar masses obtained using the formulae of Taylor et al. (2011) and Bell \& de Jong (2001). The lower right panel shows the differences in stellar masses when using the models of Bell and de Jong, but assuming two different metallicities. }
   	\label{fig:stellar_mass}
\end{figure}

\indent To have an independent test for the low mass galaxies and to assess the feasibility of Eq. 1 for the galaxies with  M$_*$ < 10$^{7.5}$ M$_{\odot}$, we compare our estimates with those of \citet{Bell2001}. We adopted the linear fit model of Bell \& de Jong that uses the stellar population models of \citet{Bruzual2003}. In this model, a Salpeter initial mass function was used\footnote{The models of \citet{Bell2001} are limited to Salpeter-like initial mass functions, which restricts us from using for example Kroupa initial mass function.} and metallicities of $\log_{10}(Z/Z_{\odot})$ = 0 and -0.4. Any lower metallicities are not covered by the Bell \& de Jong models and thus we are limited to use these rather high metallicities, although for example a dwarf galaxy with M$_*$ = 10$^{7.5}$ M$_{\odot}$ has $\log_{10}(Z/Z_{\odot})$ = -1 (see \citealp{Janz2016b}). Assuming  solar metallicity, the equation for the mass to light ratio is
\begin{equation}
\log_{10} \left( \frac{ \mathrm{M}}{\mathrm{L_R}} \right) = -0.80+1.21\times(V-I),
\end{equation}
where $V$ and $I$ are magnitudes in the Johnson V and I band filters, and $L_R$ is the luminosity in Johnson R-band. In the case of $\log_{10}(Z/Z_{\odot})$ = -0.4, the formula becomes
\begin{equation}
\log_{10} \left( \frac{ \mathrm{M}}{\mathrm{L_R}} \right) = -0.87+1.27\times(V-I).
\end{equation}
Using the transformations of \citet{Jordi2006} between the Johnson V,R,I- and the SDSS g',r',i'-filters (see Appendix E), and transforming luminosity to magnitudes we can rewrite Eq. 3 as follows:
\begin{equation}
\log_{10} \left( \frac{ \mathrm{M}_{*}}{\mathrm{M}_{\odot}}\right) = 1.04 + 0.817 \times ( g'-i') - 0.4\times M_{r'}  + 0.1\times( r'-i').
\end{equation}
While comparing Eqs. 2 and 5 (see Fig. \ref{fig:stellar_mass}), we find that the Bell and de Jong calibration is very similar to the one of Taylor et al., differing only by having a lower constant term and a slightly different dependence on the g'-i' and r'-i' colors. As the range in the colors of FDSDC galaxies is quite narrow and the color terms in Eqs. 1 and 5 are small, r'-band magnitudes work as a good proxy for the stellar mass. The effects of metallicity in the model is negligible (see the lower right panel of Fig. \ref{fig:stellar_mass}). Thus, these mass estimates give consistent results within $\approx$10\% accuracy.

\indent Due to the systematic uncertainties in the $M/L$-ratios for low luminosity galaxies, we do most of the analysis in this work using the r'-band luminosity as a proxy for the stellar mass. However, in crucial parts of the analysis, we also check that the results hold when using the estimated stellar masses instead of luminosities. For the stellar masses we adopt the values from \citet{Taylor2011} (Eq. 2). Also the stellar masses for the giant ETGs in Section 9 have been calculated in the same way.

\section{Dwarf galaxy distribution}

\subsection{Locations of the galaxies and their environment}

In Fig. \ref{fig:galaxy_locations} we show the locations of the galaxies in our sample. It is clear that most of the galaxies are strongly concentrated around the center of the Fornax cluster forming a relatively symmetrical and dense main cluster. Another clear but less massive and more loose structure is the Fornax A group southwest from the main cluster, where the galaxies are grouped around the shell galaxy NGC 1316  \citep{Iodice2017a}. The Fornax A galaxy distribution seems to be skewed toward the main cluster. The Fornax cluster is located in a filament of the cosmic web so that the filament crosses the cluster in the North-South axis (see \citealp{Nasonova2011} and Venhola et al. 2018). One would expect the infall of galaxies to be enhanced from the direction of the filament, but no clear overdensities toward those directions can be seen in our data.

\indent In order to better see the sub-structures, we convolved the galaxy distribution with a Gaussian kernel with a standard deviation of $\sigma$ = 15 arcmin. We also used mass weighting for the smoothing. The smoothed distributions with and without mass weighting are shown in Fig. \ref{fig:smoothed_density.py}.  As expected, in the smoothed images the two overdensities in the main cluster and in the Fornax A sub-group are clearly pronounced. From the non-weighted distribution it appears that the galaxy distributions are not perfectly concentrated around NGC 1399 and NGC 1316, but are slightly offset toward the space between those two giant galaxies. The distribution of the galaxies in the center of the cluster has the same two-peaked appearance as the one found by \citet{Ordenes-Briceno2018}, who analyzed the distribution of the Fornax dwarfs using the Next Generation Fornax Survey data (NGFS, \citealp{Munoz2015}). Also the giant ETGs and GCs in the cluster are concentrated to this same area (\citealp{Iodice2018},\citealp{DAbrusco2016}).

\begin{figure}
	\resizebox{\hsize}{!}{\includegraphics{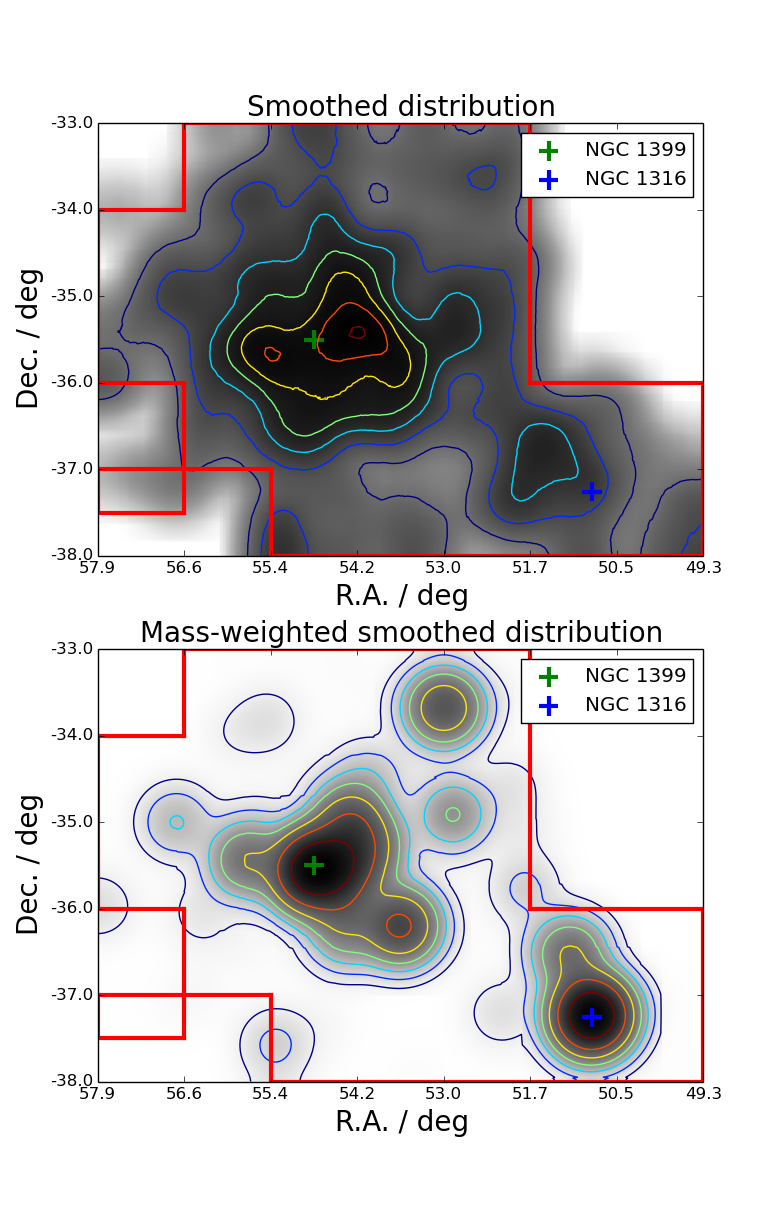}}
    \caption{Smoothed distributions of the galaxies in the Fornax cluster. The upper panel shows the surface number density distribution of galaxies using a Gaussian convolution kernel with $\sigma$ = 15 arcmin. The lower panel shows the smoothed surface mass density distribution using the same kernel. The colored contours in the upper and lower panels highlight the iso-density curves with linear and logarithmic spacing, respectively. The green and blue crosses show the locations of NGC 1399 and NGC 1316, respectively. The red lines frame the area covered by the FDS. In the figures north is up and east is toward left.}
   	\label{fig:smoothed_density.py}
\end{figure}

\indent Due to the relatively symmetric distribution of the galaxies in the Fornax cluster, it is tempting to use the cluster centric distance as a proxy for the galaxy environment and the infall time. We expect the galaxies in the outskirts of the cluster to have spent less time in the cluster environment than the ones in the inner parts. To test how well the cluster-centric distance correlates with the projected galaxy density, we calculate the number of cluster galaxies within 30 arcmin projected radius from each galaxy. We then measure this parameter for all the galaxies and in Fig. \ref{fig:dist_vs_dens} plot this parameter against the cluster-centric distance. The number of the projected neighbors drops almost monotonically toward the outer parts of the cluster. Only in the Fornax A group deviations from this relation occur. In principle, the projected galaxy density would be a sufficient measure to define the galaxy environment, but since there are some galaxy clumps that may be due to projection effects without any physical meaning, the cluster-centric distance provides a more robust measure. In summary, for our sample, a shorter cluster-centric distance can be interpreted as higher galaxy density environment. Additionally, galaxies located at short cluster-centric distances have on average entered the cluster earlier than those in the outskirts. Thus cluster-centric distance is a more fundamental parameter than the projected galaxy density.  

\indent A higher galaxy density is expected to increase the strength of harassment between galaxies, whereas higher relative velocities weaken the strength of individual tidal interactions (Eq. 8.54 of \citealp{BinneyTremaine2008} and Eq. 10 of this study). According to \citet{Drinkwater2001} the line of sight velocity dispersion of the main cluster is $\sigma_V$ = 370 km s$^{-1}$ and in the Fornax A sub-structure $\sigma_V$ = 377 km s$^{-1}$. Assuming that both structures are virialized, Drinkwater et al. estimated masses of 5$\pm$2$\times10^{13}$ M$_{\odot}$ and 6$\times10^{13}$ M$_{\odot}$, for the main cluster and Fornax A, respectively. However, they also stated that, most likely, Fornax A is not virialized and thus the virial mass estimation is only an upper limit. Assuming the same mass-to-light ratios for these two sub-clusters, Drinkwater et al. estimated the mass of the Fornax A to be 2$\times10^{13}$ M$_{\odot}$, being thus $\approx$1/3 of the mass of the main cluster. However, this approximation might still be underestimating the relative masses, since the galaxy population of the Fornax A is dominated by late-type dwarfs, which have on average younger stellar populations and thus lower mass-to-light ratio.

\begin{figure}[t]
\resizebox{\hsize}{!}{\includegraphics{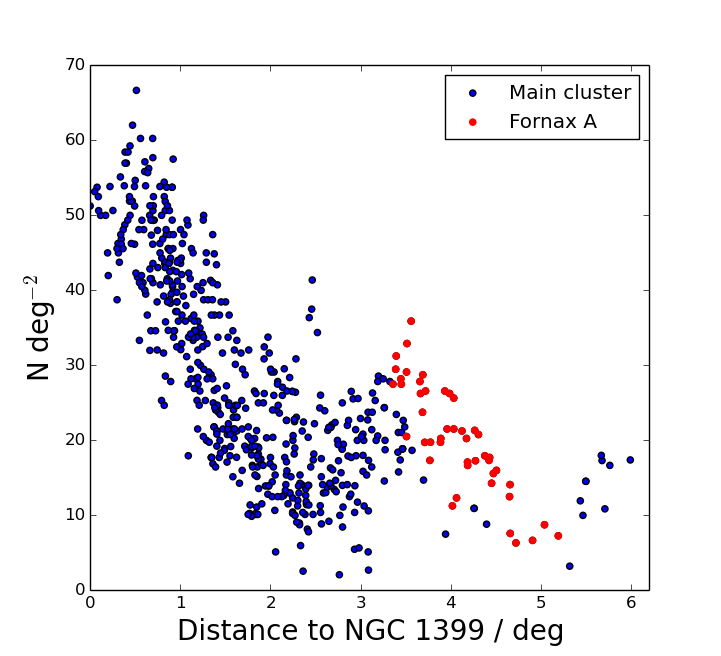}}
\caption{Galaxy surface density as a function of the distance from NGC 1399. The red points show the galaxies projected within one degree from the NGC1316, which is the central galaxy of Fornax A group. The blue points show the other galaxies.}
   	\label{fig:dist_vs_dens}
\end{figure}

\indent In addition to galaxy density, also hot X-ray emitting gas is an important factor affecting the evolution of the galaxies via ram pressure stripping. \citet{Paolillo2002} and \citet{Kim1998} studied the X-ray gas around NGC 1399 and NGC 1316, respectively, using the RÖntgen SATellite's (ROSAT) X-ray telescope's Position Sensitive Proportional Counters (PSPC) \citep{Pfeffermann1987} for imaging. In order to compare the distribution of the X-ray gas in the cluster with that of the dwarf galaxies, we took the archival ROSAT data and overlaid intensity contours on Fig. \ref{fig:galaxy_locations}. The total mass of the X-ray gas within 0.3 deg from the NGC1399 is M$_{gas}$ $\approx$ 10$^{11}$ M$_{\odot}$ \citep{Paolillo2002}, whereas the mass of the X-ray gas associated to Fornax A is only M$_{gas}$ $\approx$ 10$^{9}$ M$_{\odot}$ \citep{Kim1998}. The X-ray gas of the main cluster is detected up to radii of 150 kpc from the center, but due to the small spatial extent of the X-ray observations, in reality it might extend much further. The X-ray emitting gas is also misaligned with respect to NGC 1399. The gas is lopsided toward northwest, which is opposite to the giant ETGs in the center. Based on the X-ray gas distribution, the galaxies traveling through the center of the Fornax cluster are expected to experience much stronger ram-pressure stripping, due to larger gas-density, than the galaxies located in the Fornax A sub-group.

\subsection{Radial distribution of dwarf galaxies}

Morphological segregation is known to exist in Virgo \citep{Lisker2007} and in other clusters \citep{Binggeli1984}, and an analysis of the density-morphology relation in the Fornax cluster, and in Fornax A, gives an interesting comparison with the previous works. In particular, we expect the different environmental properties in the core and outskirts of the main cluster, and in the Fornax A sub-group, to manifest in the morphology and the physical parameters of the galaxies.

\indent In order to test whether we see different cluster-centric radial distributions among the studied morphological types, we derive radial cluster-centric surface density profiles for the late-types, dE(N)s and dE(nN)s, and compare them with the distribution of giant galaxies (M$_{r'}$ < -18.5 mag) in Fornax. We derive the morphology-density relation also with respect to Fornax A. We use circular bins centered on the central galaxies of the Fornax cluster, NGC 1399, and of the Fornax A, NGC1316. We calculate the number of galaxies in each bin and divide them by the bin area to get the surface density of the galaxies. We assume a simple Poissonian behavior for the uncertainty in counts in the bins. The partially incomplete coverage of the FDS in the outermost bins is taken into account in the radial profiles. The obtained radial surface density profiles are shown in the upper panels of Fig. \ref{fig:radial_distribution}. In the middle panels the radial cluster-centric cumulative fractions are also shown.

\begin{figure*}[t!]
	\includegraphics[width=17cm]{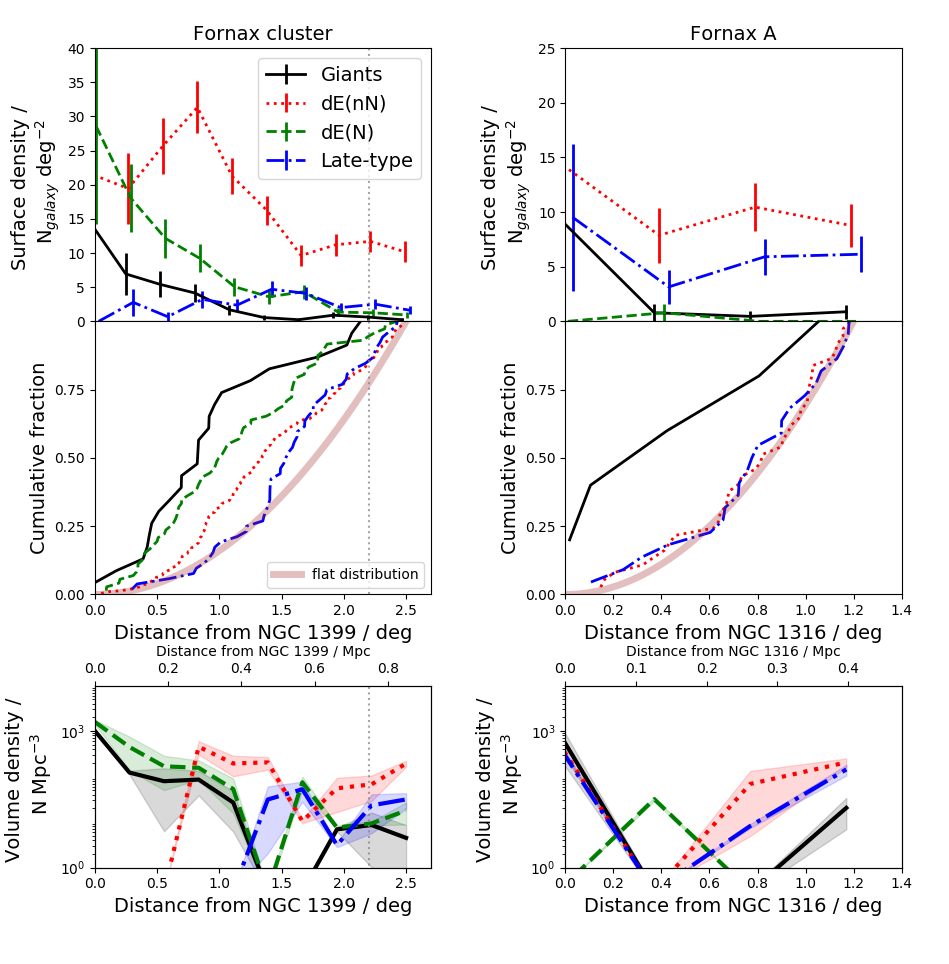}
    \caption{Upper panels show the surface number density of the galaxies as a function of the cluster-centric radius (left panels) and the distance from Fornax-A (right panels). The mid panels show the corresponding cumulative number fractions as function of radii. The black solid lines corresponds to the giant galaxies with M$_r'$ < -18.5 mag, the green dashed and red dotted lines correspond to dE(N)s and dE(nN)s, respectively, and the blue dash-dotted lines corresponds to the late-type dwarfs. For a reference, the purple opaque lines show the cumulative profile for a uniform radial surface density distribution without any cluster-centric concentration. In the bottom panels, we show the deprojected volume densities of the galaxies obtained from the radial surface density profiles using the onion peeling technique described in the Appendix B. The shaded areas correspond to the uncertainties in the deprojected profiles, estimated with Monte Carlo method. The vertical gray dotted lines in the left panels correspond to the virial radius of the cluster.}
   	\label{fig:radial_distribution}
\end{figure*}

\indent We find a clear sequence in the distributions of the different morphological types in the main cluster (upper left panel in Fig. \ref{fig:radial_distribution}): from the least centrally concentrated to the most concentrated groups are the late-type dwarfs, dE(nN)s, and dE(N)s. We use a Kolmogorov-Smirnov test to see whether these differences are statistically significant. In Table \ref{table:rad-dist-p} we show the p-values for the assumption that the observed distributions are drawn from the same underlying distribution. In this and later tests in this paper, we require p < 0.05 (>2$\sigma$ deviation from the normal distribution) for concluding that two samples differ significantly. We find that the differences are indeed significant between all the samples except between the dE(N)s and giants. 

\indent In Fornax A (upper right panel in Fig \ref{fig:radial_distribution}) the morphology-density relation is not as clear as in the main cluster. The giants are more centrally concentrated than the dwarfs, but the distributions of the early- and late-type dwarfs do not differ significantly from each other, and are only slightly concentrated toward NGC 1316 in the innermost 0.5 deg area. 

\indent To further understand the spatial distribution of the galaxies in the center of the Fornax cluster, we used the onion peeling deprojection method (described in Appendix B) for obtaining the galaxy volume densities. The deprojected distributions of the main cluster and the Fornax A sub-group are shown in the bottom row of Fig. \ref{fig:radial_distribution}. The deprojected radial distributions reveal that there are no dE(nN)s nor late-type dwarfs within the innermost 0.5 deg from the cluster center. This radius also roughly corresponds to the projected radius, which encloses the area where the giant ETGs are densely concentrated and where their outer parts are disturbed \citep{Iodice2018}. The distribution of the late-type dwarfs shows an even larger de-populated core than for the dE(nN)s, extending to r = 1.5 deg. The large core devoid of late-type galaxies and dE(nN)s implies that the environment in the cluster core causes either morphological transformations in those galaxies that  enter the core, or that the late-type dwarfs and dE(nN)s get dissolved in that part of the cluster.

\begin{table}
\caption{Results of the pairwise Kolmogorov-Smirnov tests used for comparing the radial distributions of the different types of galaxies. The p-values correspond to the hypothesis that a pair of samples is drawn from the same underlying distribution.}          
\label{table:rad-dist-p}      
\centering               
\begin{tabular}{c | c c c }     
\hline\hline            
 & Giants  & dE(nN)  & dE(N) \\
 \hline
dE(nN) & 5.5e-04 & &  \\
dE(N)  & 3.1e-01 & 6.5e-04 &  \\
dIrr   & 4.5e-05 & 5.9e-03 & 2.0e-05 \\
\hline                              
\end{tabular}
\end{table}

\indent Since our analysis of the radial profiles showed that the cluster core is almost devoid of the dE(nN) and late-types, and yet we observe them with the projected locations near the cluster center, it is useful to quantify how the projected cluster-centric 2D-distance correlates with the 3D-distance. To find the expected 3D-distance for any given 2D-distance, we integrate along the line of sight through the previously obtained volume density distributions, and calculate the density weighted mean 3D-radius of the galaxies. We show the relation between the 2D- and 3D-distances in Fig. \ref{fig:loc_likelyhood}. To quantify the width of the  distribution, we also calculate the 25\% and 75\% quantiles at the given 2D-distance.

\begin{figure}
	\resizebox{\hsize}{!}{\includegraphics{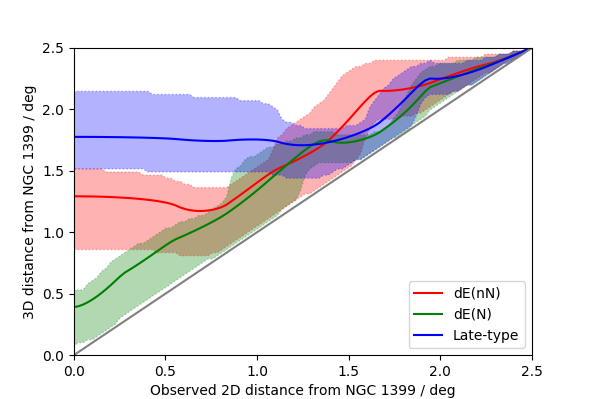}}
    \caption{Relations between the projected 2D distances and the likely 3D distances for the different morphological types. The green, red, and blue lines correspond to dE(N), dE(nN), and late-type dwarfs. The shaded areas show the range within the 25\% and 75\% in the probability distributions.}
   	\label{fig:loc_likelyhood}
\end{figure}

\indent We know that in the Virgo cluster the relative abundance of the dE(nN)s, dE(N)s, and the late type dwarfs correlates with the density of the galaxy environment \citep{Lisker2007}. To quantify the behavior in our data, we plot the early-type galaxy fraction (N$_{early}$/N$_{all}$) and the nucleation fraction (N$_{nuc}$/N$_{early}$) in the main cluster and Fornax A, as a function of the radius from the central galaxies NGC 1399 and NGC 1316, respectively (see Fig. \ref{fig:type_fractions}). From the upper left panel we see that the early-type fraction declines toward the outskirts of the cluster, being 95\% in the center and 80\% in the outer parts.  Also, the nucleation fraction of the early-types decreases from $\approx$ 0.5 in the cluster center to 0.1 in the outskirts. Neither the early-type fraction nor the nucleation fraction in the Fornax A increase significantly in the vicinity of NGC1316, and they are clearly lower than in the main cluster, being much closer to the value found in the field, where $\approx$ 50\% of the dwarfs are of late type \citep{Binggeli1990}. 

\begin{figure}
	\resizebox{\hsize}{!}{\includegraphics{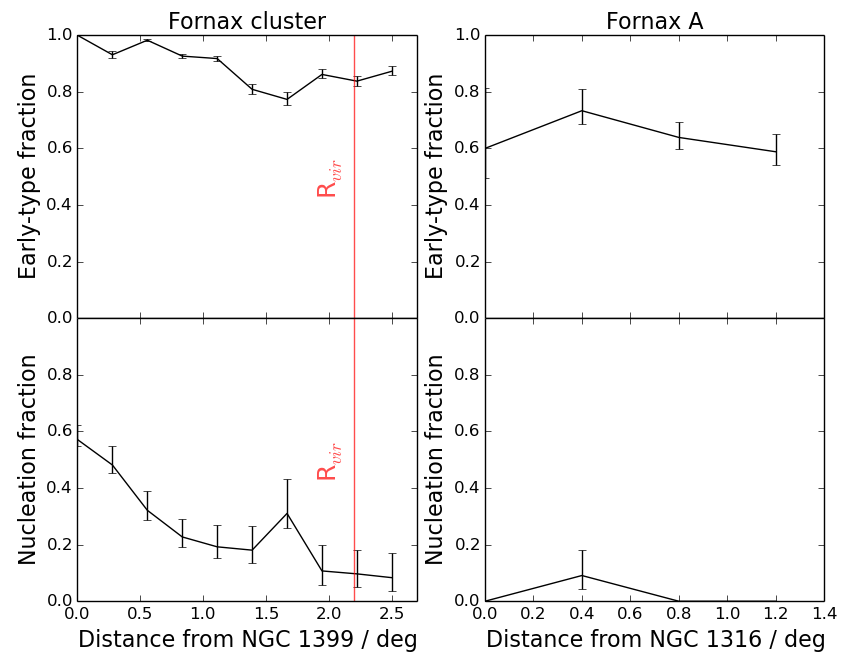}}
    \caption{Left panels show the early- type fraction (upper left) and the nucleation fraction (lower left) as a function of projected 2D distance from the central galaxy NGC 1399 of the Fornax cluster. Right panels show the same relations as a function of 2D distance from the central galaxy NGC 1316 of the Fornax A sub-group. The red horizontal lines in the left panels show the virial radius of the main cluster.}
   	\label{fig:type_fractions}
\end{figure}

\section{Luminosity function}

\subsection{In the whole cluster}
In Fig. \ref{fig:morphology_luminosity_func} (upper panel) we show the r'-band luminosity functions and the cumulative luminosity functions of the different types of dwarf galaxies. Late-type dwarfs in our sample are slightly more luminous than dE(nN)s (p=0.034), and dE(N)s are on average more luminous than the other two types (p<10$^{-5}$ for both). If we compare the luminosity functions of the early- and late-types, without considering nucleation, the luminosity functions do not differ (p=0.17).

\begin{figure}
	\resizebox{\hsize}{!}{\includegraphics{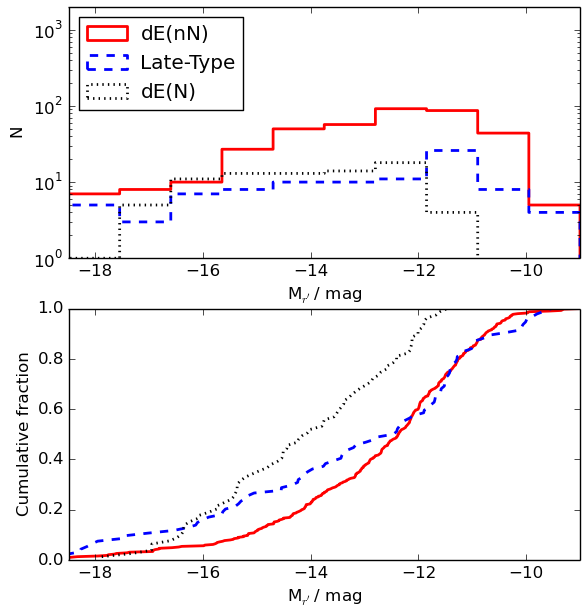}}
    \caption{Upper panel shows the r'-band luminosity function for the dE(nN) (red solid line), dE(N) (black dotted line) and late-type dwarf galaxies (blue dashed line). The lower panel shows the cumulative luminosity function of the galaxies with the corresponding colors, respectively.}
   	\label{fig:morphology_luminosity_func}
\end{figure}

\indent We also show the nucleation and early-type fraction of the Fornax cluster dwarfs as a function of galaxy magnitude (see Fig. \ref{fig:nuc_frac}). We find that the nucleation fraction peaks at M$_{r'}\approx$ -16 mag and decreases quickly toward higher and lower luminosity galaxies. This seems to be true independent of the cluster-centric distance. \citet{Ordenes-Briceno2018} analyzed the NGFS galaxy sample in the center of the Fornax cluster (\citealp{Eigenthaler2018}). They found a nucleation fraction of $\approx$90\% at M$_{r'}\approx$ -16 mag. In the same area, the nucleation fraction obtained by us agrees well with the value of \citet{Ordenes-Briceno2018}. For comparison, \citet{Cote2006} who studied the nucleation fraction in the Virgo cluster using the Hubble Space Telescope (HST) images, found a similar peak of the nucleation fraction as we found, around M$_B$ = -16.5 mag. Near the peak at least 80 \% of the galaxies have nuclei, consistent with our study. The result of \citet{Cote2006} for the Virgo dwarfs was later confirmed by \citet{SanchezJanssen2018} who identified a similar peak in the nucleation fraction, and using the deep dwarf galaxy sample of the NGVS showed that the nucleation fraction steeply decreases toward the lower luminosities. The early-type fraction (the lower panel of Fig. \ref{fig:nuc_frac}) has some dependence of the magnitude as well, so that for galaxies with M$_{r'}$ < -16 mag the early-type fraction is lower than for the less luminous galaxies. The same result holds also in stellar mass bins.
%The differences in the nucleation fractions of the dEs in the inner and outer parts of the cluster suggests that not only the mass of the galaxy, but also the galactic environment matters in the formation of the nucleus.

\begin{figure}[h]
	\resizebox{\hsize}{!}{\includegraphics{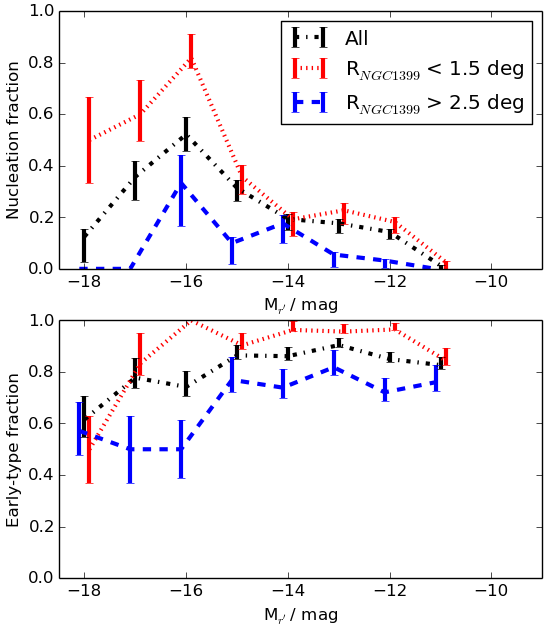}}
    \caption{ Nucleation fractions as a function of absolute r'-band magnitude at different cluster-centric radii. The red dotted, blue dashed, and black dash-dotted lines show the nucleation fraction within 1.5 deg and outside 2.5 deg from the cluster center, and for all galaxies, respectively.}
   	\label{fig:nuc_frac}
\end{figure}

\subsection{Luminosity function in radial bins}

We calculate the luminosity distributions of the galaxies in four spatial bins: at R < 1.5 deg\footnote{1.5 deg corresponds to two core radii \citep{Ferguson1989} or $\sim$450 kpc at the distance of the Fornax cluster} from NGC 1399 (Bin 1), at 1.5 deg < R < 2.5 deg (Bin 2), at R > 2.5 deg (Bin 3), and the galaxies within one degree from the Fornax A group (Bin A). In all spatial bins we make a histogram of the galaxy luminosity distribution and fit it with a Schechter function (\citealp{Schechter1976}):
\begin{equation}
n(M)\mathrm{d}M = 0.4 \ln(10) \phi^*\left[ 10^{0.4(M^*-M)} \right ]^{\alpha+1}\exp\left(-10^{0.4(M^*-M)}  \right ) \mathrm{d}M,
\end{equation}
where $\phi^*$ is a scaling factor, $M^*$ is the characteristic magnitude where the function turns, and $\alpha$ is the parameter describing the steepness of the low luminosity end of the function. We only fit the galaxies with magnitudes between -18.5 mag < M$_{r'}$ < -10.5 mag, since their parameters can be well trusted and where our sample is at least 50\% complete even in the faintest bin.

\indent The completeness of a sample is often defined by finding the surface brightness and size limits where at least 50\% of the galaxies having those parameters are included. This means that, for a given magnitude and surface brightness we have a certain probability of detecting a galaxy, which probability decreases near the completeness limits, and causes a bias to the sample. Another bias may be caused by the selection limits of the galaxy sample. As we show in the Appendix A (Fig. \ref{fig:detection_completeness}), around M$_{r'}$ = -12 -- -10 mag galaxies with very small or very large R$_e$ start to be excluded due to the detection limits. To correct for the completeness bias, a magnitude dependent completeness correction can be applied to the luminosity function. This correction thus scales up the number of galaxies especially in the lowest luminosity bins where the sample is the most incomplete. However, there is a caveat in this approach: as galaxies with a given magnitude, have a range of surface brightnesses, and the completeness drops toward lower surface brightnesses, the completeness of the galaxies is thus dependent on the galaxies' surface brightness distribution. Venhola et al. used flat R$_e$ and $\bar{\mu}_{e,r'}$ distributions within the ranges of R$_e$=100 pc -- 3 kpc and $\bar{\mu}_{e,r'}$= 22--30 mag arcsec$^{-2}$ for defining the completeness of the FDSDC detections. As there is no way to know the surface brightness distribution of the galaxies that are too faint to be detected, the only way to make the completeness correction is to assume some distribution as Venhola et al. did. Due to these uncertainties in the completeness corrections, we make the luminosity function fits both without any corrections and by correcting the luminosity bins for the completeness according to the completeness tests  by Venhola et al. (2018). As Venhola et al. showed, the FDSFC sample misses some known LSB galaxies in the Fornax cluster, and therefore the real slope of the luminosity function is likely to be somewhere between the values found for the completeness corrected and non-corrected luminosity functions. 

\indent In Fig. \ref{fig:luminosity_function} (upper panels) we show the luminosity functions and the corresponding Schechter fits in the four spatial bins. Shown separately are also the luminosity functions for the different morphological types (lower panels). From the raw data we obtain the faint end slopes of $\alpha$ = -1.14$\pm$0.10, -1.13$\pm$0.06, -1.25$\pm$0.13 and -1.44$\pm$0.13 in the bins from the innermost to the outermost, and with the completeness corrections $\alpha$ = -1.31$\pm$0.07, -1.27$\pm$0.05, -1.38$\pm$0.11 and -1.58$\pm$0.16, respectively. The Schechter fits for the dwarf galaxies give similar faint end slopes in the three bins located in the main cluster area. Apparently, Fornax A differs from the other bins, showing a power-law like distribution. The difference can be also clearly seen from the cumulative luminosity functions shown in Fig. \ref{fig:luminosity_function_cum}. The peculiar luminosity function of Fornax A can be related to the massive stellar streams around NGC 1316 (see \citealp{Iodice2017a}), of which some may originate from disrupted dwarf galaxies. However, due to the small number of galaxies we cannot robustly exclude the possibility that the underlying distribution in Bin 4 is similar to the other bins (see Table \ref{table:lumi-p}), and we do not see the exponential part of the function due to the low number of galaxies. The luminosity functions of different morphological types look qualitatively similar in all four bins, except that the number of dE(N)s decreases toward the outskirts of the cluster.

\indent In the upper left panel of Fig. \ref{fig:luminosity_function}, we also show (with the red line) the luminosity function after removing the late-type dwarfs that are likely to appear in the central parts of the cluster only due to the projection effects. This correction further flattens the faint end slope of the luminosity function, compared to the completeness corrected value.

\indent The faint end slope of the luminosity function in the center of the Fornax cluster have been studied by \citet{Hilker2003} and \citep{Mieske2007}. Both of those studies apply a completeness correction and find a slope of $\alpha$ = -1.1$\pm$0.1, which is consistent with the faint end slope that we find in the center of the cluster when the completeness correction is not applied or when the late-type dwarfs are excluded. Our completeness corrected value of $\alpha$ is slightly steeper than theirs.

\indent  For comparison, \citet{Ferrarese2016} who measured the faint end slope in the core of the Virgo cluster using the NGVS galaxy sample, find a completeness corrected $\alpha$ = -1.34$^{+0.017}_{-0.016}$, which is in good agreement with our value (-1.31$\pm$0.07). For the NGVS the luminosity function in the outer parts of the Virgo cluster has not been studied. Other works, including also the outer parts of the Virgo cluster, have shallower observations than ours, and there is no clear consensus about the slope in the outer parts (see the discussion of \citealp{Ferrarese2016}). In the Coma cluster, which is an order of magnitude more massive than Virgo and Fornax clusters, \citet{Bernstein1995} measure  $\alpha$ =-1.42$\pm$0.05, which is consistent within the errors with the values obtained by Ferrarese et al. in the Virgo and by us in the Fornax cluster. \citet{Ferrarese2016} also study the luminosity function in the Local Group using similar sample selection criteria as they used in Virgo, but without applying the completeness correction, and obtain a faint end slope of $\alpha$ = -1.22$\pm$0.04. The obtained value is slightly shallower than in the clusters. An additional study that has a similar depth as the above mentioned ones is by \citet{Trentham2002}, who analyze the mean faint-end slope (-18 mag < M$_R$ < -10 mag) of the luminosity functions in five different groups and clusters including Virgo and Coma clusters. Like us, they also made the membership classifications for the faint-end galaxies based on their optical morphology, but did not apply any completeness correction for the luminosity function fitting. They found a mean slope of $\alpha$ = -1.2. Interestingly, the differences between the faint-end slopes of the luminosity functions in these different environments are smaller than the difference between the completeness corrected and non-corrected $\alpha$-values in the Fornax cluster, which indicate that these differences might emerge artificially from the correction.

\indent The faint end slopes of the early-type galaxy luminosity functions in Hydra I, $\alpha$ = 1.13 $\pm$ 0.04, and Centaurus clusters, $\alpha$ = 1.08 $\pm$ 0.03, were studied by \citet{Misgeld2008} and \citet{Misgeld2009}, respectively. In these two works the luminosity functions have been corrected the same way as \citet{Hilker2003} did in the Fornax cluster. The faint-end slopes that they find are similar to the value we found for the early-type dwarfs in the center of the Fornax cluster, $\alpha$ = 1.09 $\pm$ 0.10.

\indent From our data we conclude that the luminosity function does not change significantly in different parts of the Fornax cluster nor between the compared environments. A caveat in our analysis of the Fornax cluster is that our imaging extends only slightly outside the virial radius. To further consolidate the universality of the luminosity function, one should have a control sample further away from the Fornax cluster center.

\begin{figure*}[!ht]
	\includegraphics[width=17cm]{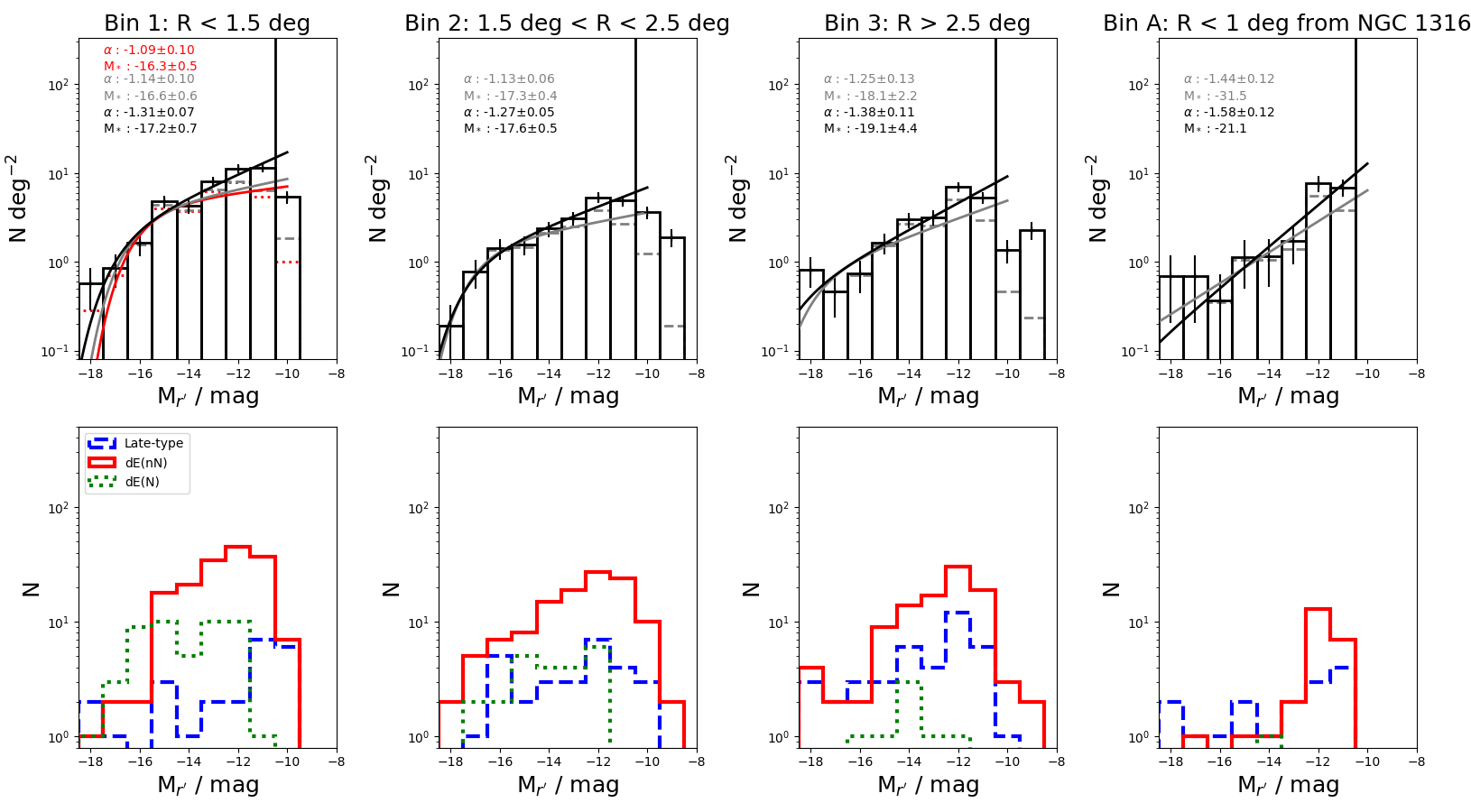}
    \caption{Four upper panels show the luminosity function in the different spatial bins around the Fornax main cluster and the Fornax A group. At the distance of the Fornax cluster one degree corresponds to 350 kpc. The gray dashed histograms show the observed number of galaxies, and the black histograms show the numbers after correcting for the completeness. The black vertical lines indicate the 50\% completeness limit of our catalog. The gray and black curves shows the fitted Schechter function for the observed and completeness corrected luminosity functions, respectively, and the red line in the upper left panel shows the fit for the innermost bin, when the late-type galaxies are excluded from the fit. The fit parameters of each bin are shown in the upper left corner of the panels with the colors corresponding to the colors of the fitted functions. The lower panels show the luminosity distribution for the different types of galaxies in the bins.}
    \label{fig:luminosity_function}
\end{figure*}

\begin{figure}
		\resizebox{\hsize}{!}{\includegraphics{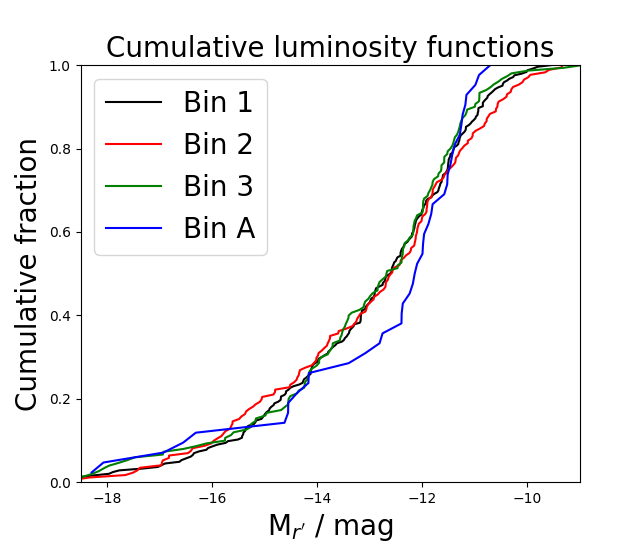}}
    \caption{Cumulative luminosity functions in the bins defined in the Fig. \ref{fig:luminosity_function}. }
   	\label{fig:luminosity_function_cum}
\end{figure}

\begin{table}
\caption{Kolmogorov-Smirnov-test p-values for the hypothesis that given two luminosity distributions are drawn from the same underlying luminosity function. The bins correspond to different cluster-centric distances so that Bin 1 includes galaxies within R < 1.5 deg from NGC 1399, Bin 2: 1.5 deg < R < 2.5 deg, and Bin 3: R > 2.5 deg. Bin 4 includes the galaxies within one degree from the NGC 1399.}          
\label{table:lumi-p}      
\centering               
\begin{tabular}{c c c c}     
\hline\hline            
 & Bin 1  & Bin 2  & Bin 3 \\
 \hline
Bin 2 & 0.83 & &   \\
Bin 3 & 0.98 & 0.50 &   \\
Bin 4 & 0.11 & 0.229 & 0.31 \\

\hline                              
\end{tabular}
\end{table}

\section{Dwarf galaxy structure}

\subsection{Structural scaling relations}
In Fig \ref{fig:mag-par-relations} we show the measured structural parameters of the galaxies in our sample as a function of their total absolute r'-band magnitude M$_{r'}$. Both for the late- and early-type galaxies the parameters show correlations with M$_{r'}$. In the size-magnitude plane (top panels in Fig. \ref{fig:mag-par-relations}) the late-type galaxies show almost linear relation between M$_{r'}$ and $\log_{10}\mathrm{R}_e$, whereas the early-type galaxies show a knee-like behavior  so that while the galaxies with -18 mag < M$_{r'}$ < -14 mag have nearly constant R$_e$, the galaxies fainter than that become smaller. One should keep in mind that the largest galaxies among those fainter than M$_{r'}$ = -14 mag might not be detected\footnote{We will study those LSB galaxies in more detail, in a future paper (Venhola et al., in preparation), in which we use a new algorithm to detect those galaxies.} because of their low $\bar{\mu}_{e,r'}$. Thus the knee might be partially reflecting the completeness limits of our sample. However, the increased number of small dwarfs (R$_e$ < 0.5 kpc) among the low luminosity galaxies is not due to any selection effect. We do not find significant differences between the scaling relations of dE(N) and dE(nN). The different relations between M$_{r'}$ and R$_e$ of the early- and late-type galaxies can also be seen in the mean effective surface brightnesses $\bar{\mu}_{e,r'}$. The late-type dwarfs with M$_{r'}$> -14 mag have brighter $\bar{\mu}_{e,r'}$ than the corresponding early-types, but for the galaxies brighter than that, the order changes.

\begin{figure*}[t]
	\includegraphics[width=17cm]{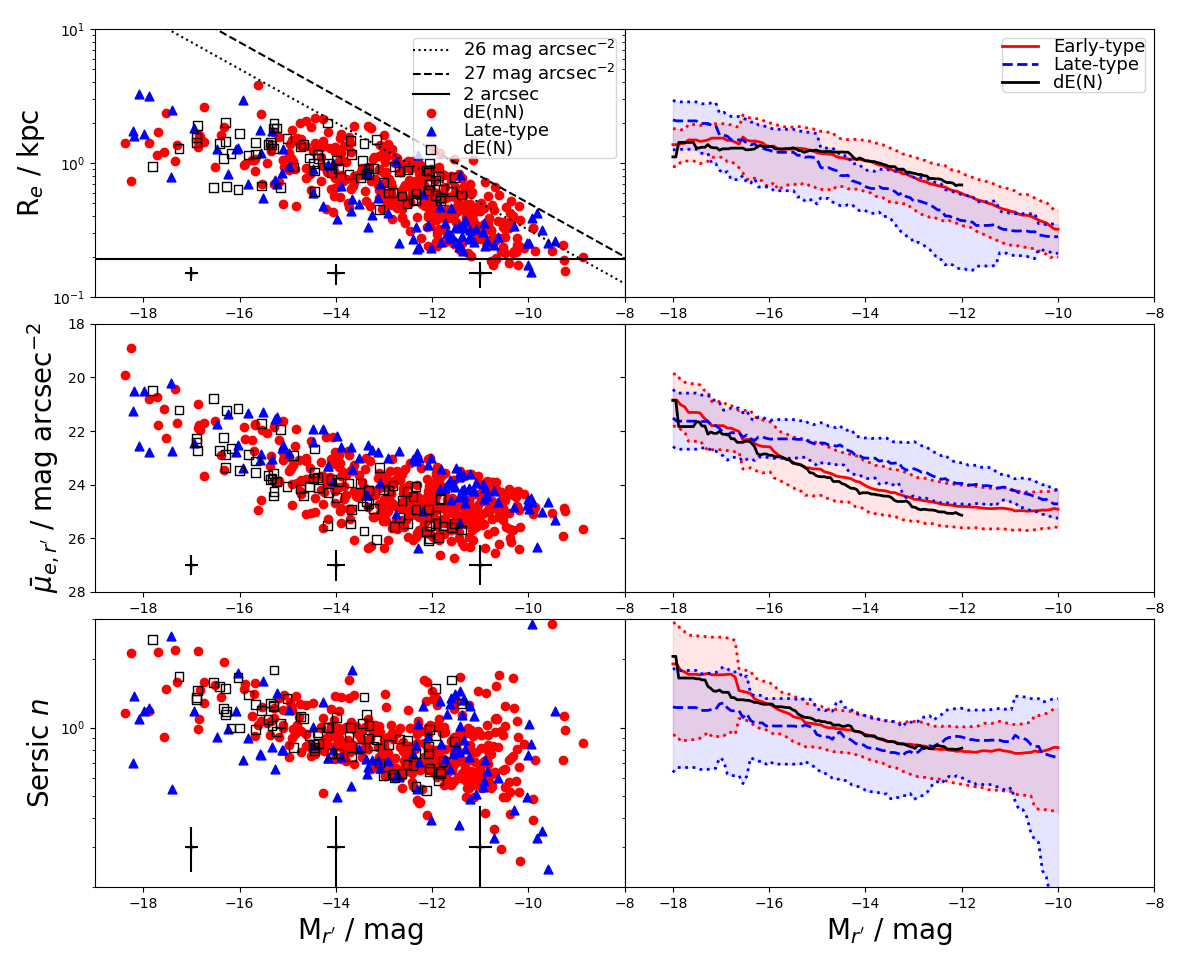}
    \caption{From top to bottom, the panel rows show effective radii (R$_e$), mean effective surface brightness in r'-band ($\bar{\mu}_{e,r'}$), and S\'ersic indices ($n$) for our sample of galaxies as a function of their total r'-band absolute magnitude (M$_{r'}$). The left panels show the individual galaxies, and the right panels show the running means of three different types using a filter size of $\Delta$M$_{r'}$ = 0.5 mag. The blue, red, and black symbols and lines correspond to late-type dwarfs, dE(nN)s, and dE(N)s, respectively. The shaded areas in the right panels correspond to the standard deviations of the points using the same running filter as for the means. For readability, we do not show the standard deviations of the dE(N)s, but they are similar to the ones of dE(nN)s. The black crosses in the left panels show the typical errorbars associated to the values at given magnitudes.}
   	\label{fig:mag-par-relations}
\end{figure*}

\indent The bottom panels of the Fig. \ref{fig:mag-par-relations} show the S\'ersic $n$ parameters of the galaxies as a function of M$_{r'}$. S\'ersic $n$ describes the shape of the galaxy's radial intensity profile so that $n$ = 1 corresponds to an exponential profile, whereas larger values mean increased peakedness of the center. The values less than one correspond to a flatter profile in the center. The S\'ersic indices are taken from the GALFIT models, which take into account the PSF convolution, and thus prevent artificial flattening of the galaxy centers for small and peaked galaxies. As mentioned before, the early-type galaxies in the magnitude range between -14 mag > M$_{r'}$ > -18 mag have nearly constant R$_e$. However, for these morphological types the S\'ersic index in this magnitude range grows from $n\approx1$ to $n\approx2$, which means that the brighter galaxies become more centrally concentrated. A similar trends between S\'ersic index and absolute magnitudes of the early-type dwarfs have also been found in Hydra I \citep{Misgeld2008} and Centaurus \citep{Misgeld2009} clusters. The fainter galaxies have nearly constant S\'ersic $n$. For the late-types, the S\'ersic $n$ grows slower than for the dEs, from slightly below one to slightly above one with increasing luminosity. As the M$_{r'}$-$n$ relations are different for the two types, the S\'ersic $n$ of M$_{r'}$ < -15 mag galaxies are higher for dEs than late-types, whereas for the galaxies fainter than that they are similar. 

\subsection{Axis-ratios}

\indent The axis-ratios of the FDSDC galaxies were obtained from the 2D S\'ersic profile fits made using GALFIT. In Fig. \ref{fig:pas} we show the apparent minor-to-major axis ratio $b/a$ distributions of the different types of dwarf galaxies in our sample. It is clear that the $b/a$ distributions move closer to unity in a sequence from late-types to dE(nN)s and dE(N)s. Qualitatively this can be interpreted as galaxies becoming rounder.

\begin{figure}
		\resizebox{\hsize}{!}{\includegraphics{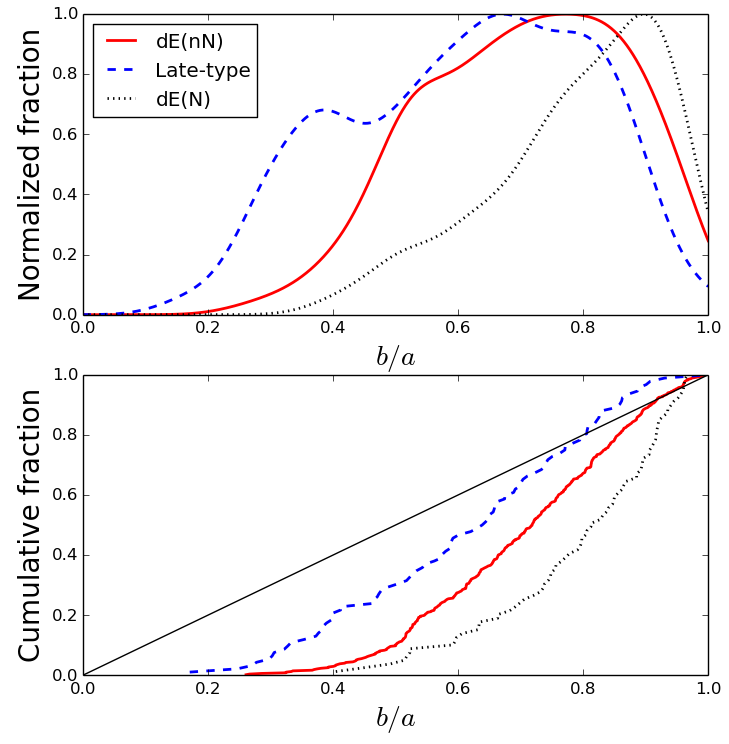}}
    \caption{Axis ratio distributions of the dE(nN)s, dE(N)s, and late-type dwarfs are shown with the red, black and blue histograms, respectively. The upper panel shows the normalized distributions and the lower panel shows the cumulative ones. The diagonal line in the lower panel represents a flat $b/a$ distribution.}
   	\label{fig:pas}
\end{figure}

\indent In order to more accurately quantify the shape distributions, we need to take into account the fact that we observe a 2D projection of the intrinsic 3D shape distribution of the galaxies. We use the same deprojection method as \citet{Lisker2007}(described in detail in Appendix C) for obtaining the intrinsic minor-to-major axis ($c/a = q$. In order to do the deprojection, we need to assume that the whole galaxy distribution consists of either prolate or oblate spheroids. In the middle panels of Fig. \ref{fig:ba_analysis}, we show the $q$ distributions  for both prolate and oblate cases. 

\begin{figure*}
	\includegraphics[width=17cm]{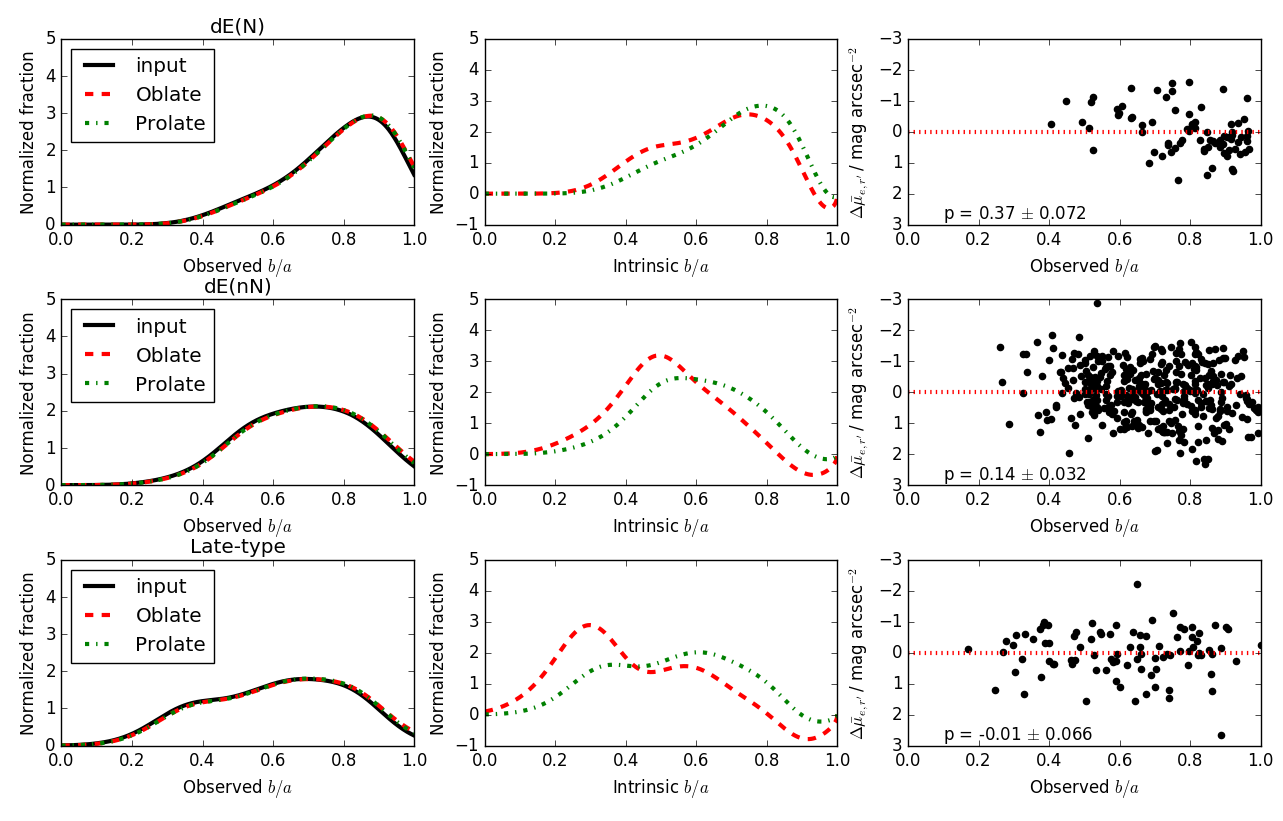}
    \caption{Intrinsic shape distributions deduced from the observed $b/a$ distributions. The left panels show the observed $b/a$ distribution with the black lines, and the reprojected $b/a$ distributions obtained from the modeled intrinsic shape distributions with the red dashed and the green dash-dotted lines for an oblate and prolate models, respectively. The middle panels show the intrinsic minor-to-major axis ($q$) distributions obtained via modeling for the oblate and prolate models. The right panels show the relation of the difference from the magnitude-surface brightness relation with the observed axis ratio. The Spearman's rank correlation coefficient p and its uncertainty are reported in each scatter plot.  The rows from top to bottom show the analysis for dE(N), dE(nN) and late-type dwarfs, respectively.}
   	\label{fig:ba_analysis}
\end{figure*}

\indent To judge whether the galaxies are better described by prolate or oblate spheroids, we can study the relation between their $b/a$ and differential surface brightness $\Delta\bar{\mu}_{e,r'}$. By $\Delta\bar{\mu}_{e,r'}$ we mean the difference in the surface brightness with respect to the mean surface brightness of galaxies with the same morphological type and luminosity. For oblate spheroids,  $\Delta\bar{\mu}_{e,r'}$ brightens with decreasing $b/a$, since when $b/a$=1 we see the system through its shortest axis. For prolate systems there should be an anticorrelation between those quantities since when $b/a$=1, we look through the longest axis of the system. The relations between $b/a$ and  $\Delta\bar{\mu}_{e,r'}$ are shown in the right panels of the Fig. \ref{fig:ba_analysis}. We also tested the statistical significance of these correlations using Spearman's rank correlation test that is robust for outliers. We find that for the dEs, the trends behave as expected for oblate spheroids, but for late-types there  is no correlation with $b/a$ and  $\Delta\bar{\mu}_{e,r'}$. The lack of correlation for the late-types can be explained if they have a significant amount of dust in their disks, which attenuates the light when observed edge-on. We argue that this is likely the case, since late-types, by definition, are gas-rich, star-forming systems, which are typically disks.

\indent Assuming that all the dwarfs are oblate spheroids, the mode of $q$ grows from 0.3 of the late-types, to 0.5 of dE(nN)s and  0.8 of dE(N)s. This means that the disks of the galaxies become thicker in the same order.Taking into account that the dEs appear deeper in the cluster than the late-types (Section 3.2), the disks of the galaxies become thicker toward the inner parts of the cluster. Since the nucleus and S\'ersic components are fitted separately for the dE(N)s, the nucleus does not bias the $b/a$ distribution of those galaxies.

\indent In the Virgo cluster, the $b/a$-distributions were analyzed by \citet{Lisker2007} and \citet{Sanchez-Janssen2016}. \citet{Lisker2007} studied the $b/a$ distributions of the M$_B$<-13 mag early-type dwarfs. They analyzed dE(N)s and dE(nN)s similarly to us, but they divided their galaxies into bright and faint dEs with the division at M$_{r}$=-15.3 mag. Since most of the dE(nN)s in our sample are fainter than that magnitude limit and most dE(N)s are brighter, it makes sense to compare our $b/a$-distributions with the "dE(N) bright" and "dE(nN) faint" samples of Lisker et al. The distribution of $q$ of dE(N)s in the Fornax cluster shows a similar potentially double-peaked distribution as the one found by Lisker et al.. A small difference between the distributions is that, in the Fornax sample, the galaxies are more abundant at the peak located at $q$=0.8, whereas in Virgo the more dominant peak is at $q$=0.5. Also the dE(nN) distributions are very similar in the two clusters, both  of them showing a clear a peak at $q$=0.5. \citet{Sanchez-Janssen2016}, who also study the $b/a$ distribution of the dwarf galaxies projected in the center of the Virgo cluster, did not divide dEs into different morphological sub-groups. Their sample is complete down to M$_{g'}$ > -10 mag. Their $b/a$ distribution is qualitatively very similar to the one of dE(nN)s in our sample. As the dE(nN)s are the most abundant type of galaxies in our sample, we conclude that there are no significant differences between the isophotal shapes of dEs in the Virgo and Fornax clusters.

\subsection{Effects of the environment on the dwarf galaxy structure}

If the environment plays an important role in galaxy evolution we should see the galaxies of the same luminosity to have different structures and/or stellar populations depending on their locations in the cluster. In Fig. \ref{fig:distance_vs_magnitude}, we divide our sample into four magnitude bins, and study how the galaxy properties vary as a function of their cluster-centric distance. We study the radial trends in the same three cluster-centric bins as in Fig. \ref{fig:luminosity_function}. According to Section 3.2, the late type galaxies with projected cluster-centric distances R<1.5 deg are likely to have 3D distance  larger than 1.5 deg, and the dE(nN)s projected within the inner R<0.8 deg, are likely to have 3D distance larger than 0.8 deg. Thus, we perform the cluster-centric trend analysis also by excluding those galaxies. In Fig. \ref{fig:distance_vs_magnitude}, we show the radial trends for all galaxies (purple lines) and for the galaxies remaining after excluding the galaxies from the inner parts of the cluster that are likely to appear there only due to the projection effects (the green lines). We also test the significance of all the correlations using the Spearman's rank correlation test, and show the correlation coefficients in the panels of Fig. \ref{fig:distance_vs_magnitude}. We base our analysis on the values that are obtained by removing the likely projection bias (green lines). All the bin means and the corresponding uncertainties are tabulated in Appendix F in Table \ref{table:tab_fig15}.

\indent In Fig. \ref{fig:distance_vs_magnitude}, we find that some of the parameters behave similarly with respect to the cluster-centric distance in all the luminosity bins, and some parameters have different trends for the high (M$_{r'}$<-16.5) and low luminosity (M$_{r'}$>-16.5) galaxies. For all the mass bins, the mean g'-r' colors of the galaxies become redder and the Residual Flux Fraction ($RFF$) decreases when going inward, meaning that the galaxies in the center are smoother than the galaxies in the outer parts. For the low luminosity dwarfs, R$_e$ becomes larger, and $\bar{\mu}_{e,r'}$ systematically fainter from the outer parts toward the cluster center, meaning that the galaxies become more diffuse. For the bright dwarfs R$_e$ becomes smaller, S\'ersic $n$ larger, and $\bar{\mu}_{e,r'}$ stays constant from the outer parts toward the cluster center. An important observation is that these cluster-centric trends are dependent on the galaxy mass.

\indent \citet{Janz2016} found that the low-mass galaxies outside the clusters are larger than the similar mass galaxies in the Virgo cluster. If we assume that galaxies gradually decrease in size, not only while entering the cluster, but also within the cluster, becoming smaller toward higher density regions, we would expect to see the same trend in our study. This trend seems to appear among the brightest dwarf galaxies in our sample, but not among the fainter ones that comprise the majority of the galaxies. As the smallest galaxies of Janz et al. have stellar masses of a few times 10$^{7}$ M$_{\odot}$ (similar to our bright dwarfs), our results for the bright dwarfs are consistent with the results of Janz et al. in the mass range they studied.

\begin{figure*}[ht]
    \centering
        \includegraphics[width=17cm]{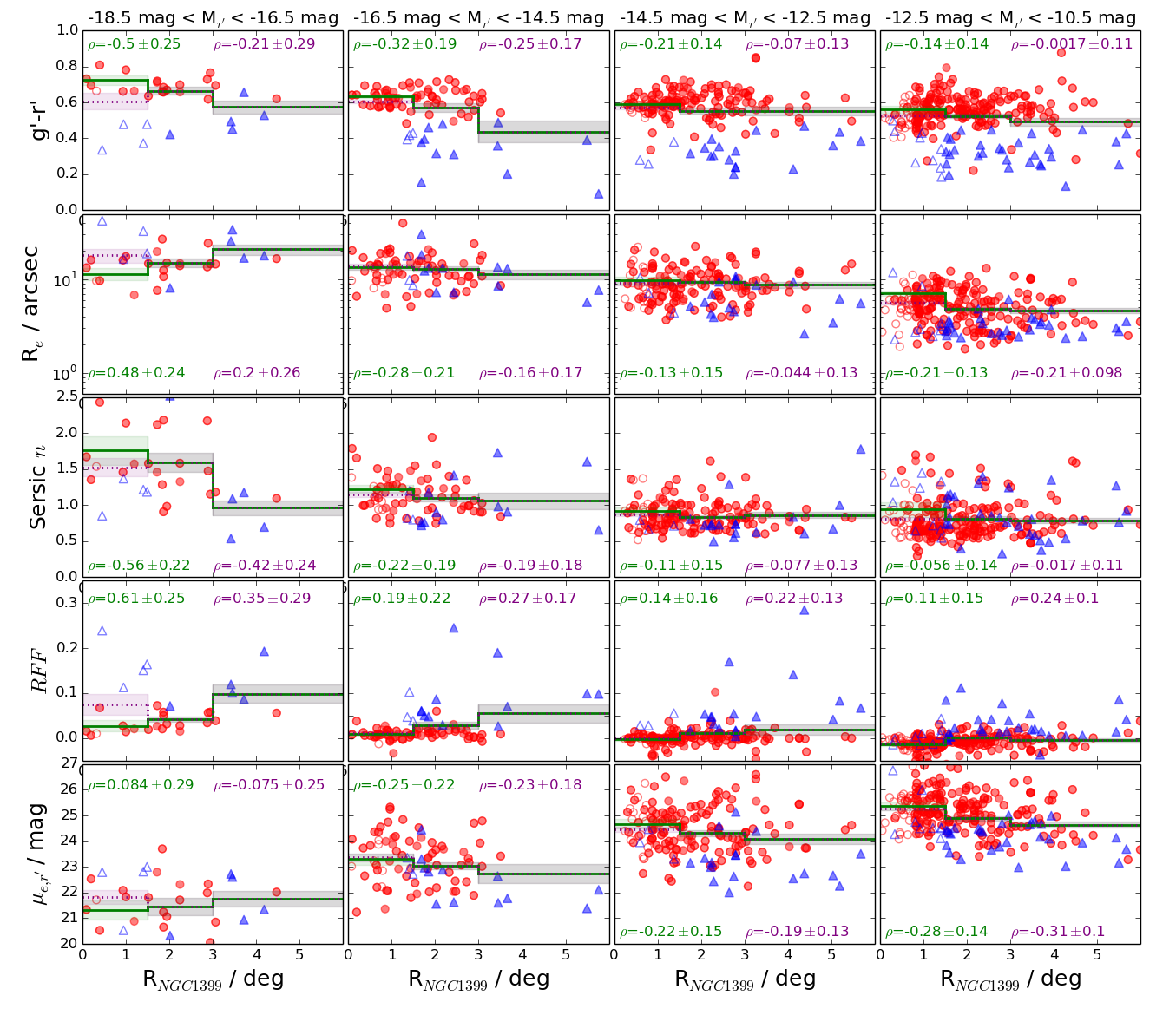}
    \caption{Panel rows from top to bottom show the g'-r' color, effective radius ($R_e$), S\'ersic index ($n$), Residual Flux Fraction ($RFF$), and the mean effective surface brightness ($\bar{\mu}_{e,r'}$) of the Fornax dwarf galaxies as a function of their cluster-centric distance. The columns correspond to different luminosity bins. The lines and shaded areas show the bin means and uncertainties of the mean for the parameters. The green and purple colors correspond to the values when the projection effects are taken in account (dE(nN)s and late-types excluded from the inner parts as found in Section 3.2), and when raw data is used, respectively. The late-type and early-type dwarfs are shown with the blue and red symbols, respectively. The unfilled symbols correspond to the galaxies excluded from the inner parts in order to minimize the projection bias. The $\rho$ parameter indicates the Spearman's correlation coefficient for the correlation of the parameter with distance, and its uncertainty.}
    \label{fig:distance_vs_magnitude}
\end{figure*}

\section{Colors of the dwarf galaxies}

\subsection{Color-magnitude relation}

In Fig. \ref{fig:cm_relations} we show the color-magnitude (CM) relations of the galaxies in the Fornax cluster. We show separately dE(nN), dE(N) and late-type galaxies. It is obvious that the morphologically selected late-type galaxies are bluer than early-types in all the CM relations. In addition to their offsets, the relations of the early- and late-type galaxies show also different slopes. The CM relations of the late-type galaxies are rather flat, so that there is not much correlation with the magnitude, whereas the early-type galaxies become bluer with decreasing total luminosity in all colors. Since the early-type galaxies become bluer toward lower luminosity, the late- and early-type galaxies have almost similar colors in the low luminosity end. This phenomenon is particulary visible in the colors including the u'-band. The dE(N)s seem to follow the same CM relation as the dE(nN)s.

\begin{figure*}[ht]
    \centering
        \includegraphics[width=17cm]{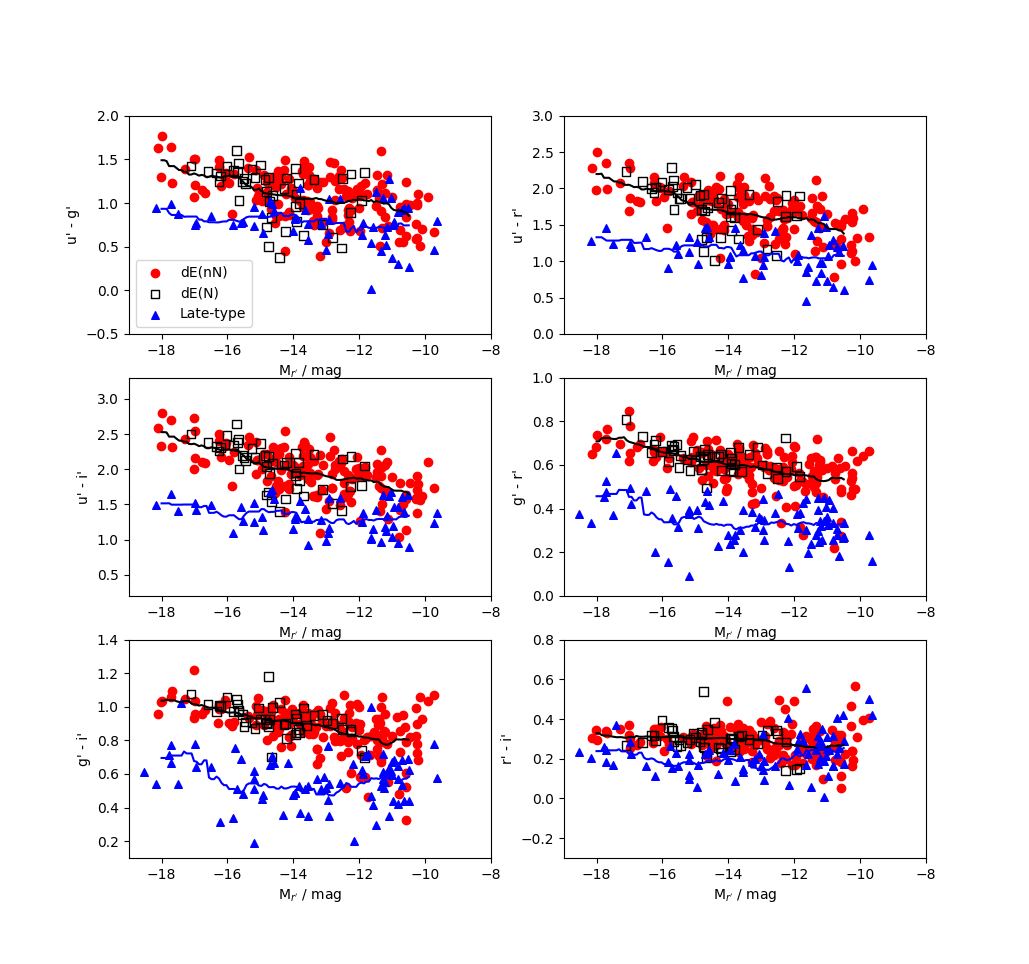}
        \caption{Color-magnitude relations of the galaxies in the Fornax cluster. The blue and red symbols correspond to the late-type galaxies and dE(nN)s, respectively, and the black squares correspond to the dE(N)s. The black and blue lines show the 2$\sigma$-clipped running means for late-types and dEs, respectively, with a kernel width of 1.6 mag.}
\label{fig:cm_relations}
\end{figure*}

\subsection{Dependence on the environment}

We have already shown that the early-type galaxies are redder than the late-type systems, and that their fraction increases toward the inner parts of the cluster. This makes the average color of the galaxies in the center of the cluster to be redder than in the outer parts. Here we concentrate on the early-type galaxies in order to see if we see possible environmental dependency also on their colors separately (see Fig. \ref{fig:cm_relations_var}).

\indent To test this, we divided our early-type galaxy sample into three spatial bins and measured their CM relations using a running mean in each bin. We measured separately the galaxies within two core radii, within the virial radius, and outside that. The CM relations in the different spatial bins, and the CM relations in the core of the Virgo cluster \citep{Roediger2017} for comparison, are shown in Fig. \ref{fig:cm_relations_var}. We transformed the MegaCAM magnitudes used by Roediger et al. into the SDSS ones by using the formulas provided at the MegaCAM web pages
(\url{http://www1.cadc-ccda.hia-iha.nrc-cnrc.gc.ca/community/
CFHTLS-SG/docs/extra/filters.html}). We find that when compared with \citet{Roediger2017}, the CM relation in the inner bin of the Fornax cluster is in good agreement with the galaxy colors in Virgo in all the colors except in g'-i, where there is an offset of $\approx$0.1 mag, between -14 mag <  M$_{g'}$ < -10 mag. As there is some uncertainty related to the filter transformations, we do not want to over-interpret the physical meaning of the offset. Within the accuracy of our measurements, we find that the CM relations of the early-type galaxies are similar in the cores of the Virgo and the Fornax clusters.

\indent When comparing the CM relations within the Fornax cluster, we find that for a given magnitude the u'-g', u'-r', and u'-i' colors become redder toward the inner parts of the cluster, whereas for g'-r' no significant differences are seen. Since the u'-band covers a wavelength range on the blue side of the Balmer jump at 4000 \AA, it is very sensitive to ongoing or recent star formation. It seems that, regardless of the fact whether the low-mass galaxies in the outer parts of the cluster are classified as dEs or not, their star formation seems to have been stopped more recently than the star formation of those galaxies in the inner parts of the cluster. Alternatively, the bluer colors of outer galaxies might be explained by them having a slightly lower metallicity than the galaxies in the inner parts. Furthermore, this color difference might be intrinsically even more significant than we observe, since as shown in Section 3, we see the galaxies in projection, which causes some of the galaxies apparently located in the cluster center to actually in the outskirts of the cluster.

\indent Radial color trends in the Fornax cluster were also found for the giant ETGs by \citet{Iodice2018}. They showed that the giant ETGs that are located within projected distance of $\approx$ 1 deg from the cluster center are systematically redder than the galaxies outside that radius. This trend is thus similar as the dwarfs have, but the color difference of the giant ETGs is also clear in the g'-r' and g'-i' colors, which is not the case for the dwarfs.

\begin{figure*}[ht]
    \centering
        \includegraphics[width=17cm]{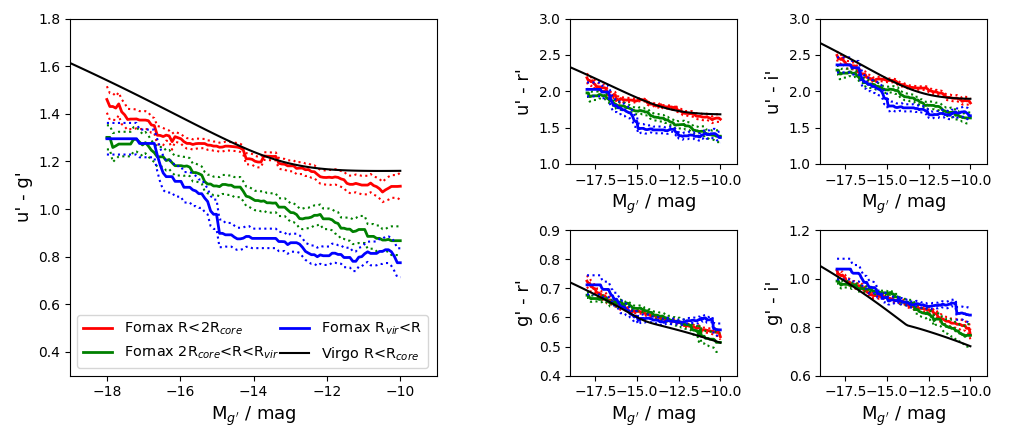}
    \caption{Panels show the running means of the color-magnitude relations in different colors for the dEs. The red, green, and blue lines corresponds to inner, middle and outer bins in the Fornax cluster, respectively. The black lines show the \citet{Roediger2017} color-magnitude relation in the core of the Virgo cluster. We use the g'-band absolute magnitudes on the x-axis instead of r'-band, since the analysis for the Virgo was done using g'-band.}
    	\label{fig:cm_relations_var}
\end{figure*}

\section{Discussion}

In the previous sections we have extensively studied the structures, colors, and distributions of dwarf galaxies in the Fornax cluster. When similar observations are available in the literature, we have also compared our measurements with those. In the following sub-sections we discuss how these observations can be understood in the context of galaxy evolution.

\subsection{Trends in the dwarf galaxy populations as a function of cluster-centric distance}

We studied how the properties of the dwarf galaxies in the Fornax cluster change as a function of the cluster-centric distance, and found the following changes in their structure and colors: 
\begin{itemize}
\item For all the galaxies, their colors become redder (Sections 5.3 and 6.2, Fig. \ref{fig:distance_vs_magnitude} and Fig. \ref{fig:cm_relations_var}) and $RFF$ decreases when moving inward that is they become more regular.

\item The low-luminosity dwarf galaxies (M$_{r'}$ > -16 mag) become more diffuse toward the inner parts of the cluster (Fig. \ref{fig:distance_vs_magnitude}).

\item For the brightest dwarfs, their $\bar{\mu}_{e,r'}$ stays constant, S\'ersic $n$ increases and R$_e$ decreases toward the inner parts of the cluster Fig. \ref{fig:distance_vs_magnitude}). 

\item The shape of the galaxies become rounder (Section 5.2) in a sequence from late-types, to dE(nN)s to dE(N)s. As the galaxies are also mostly located at different cluster-centric radii in that same order, the mean cluster-centric distance decreasing from late types to dE(N)s (Section 3.2), we can conclude that the shapes of the galaxies become rounder toward the inner parts of the cluster.
\end{itemize}

We found that the luminosity function remains constant in the different parts of the cluster (Section 4.2), but the fractions of the relative contributions of the morphological groups change. The least centrally concentrated are late-type dwarfs, then dE(nN), and dE(N) (Section 3.2). When taking into account the projection effects, we found that there are probably no late-type galaxies within 1.5 deg from the cluster center, and no non-nucleated dEs within the innermost 0.8 deg of the cluster. This innermost area of the cluster is densely populated by giant ETGs (\citealp{Iodice2018}) and intra-cluster X-ray gas (\citealp{Paolillo2002}), and in this area also asymmetric outer parts have been detected in the giant ETGs (\citealp{Iodice2018}), indicating strong tidal forces in that area. The ETGs in the center of the cluster are also systematically redder than the giant ETGs outside that dense core, indicating that their star formation was quenched early on (\citealp{Iodice2018}).  These observations, combined with the structural trends strongly suggest that the cluster environment is transforming the dwarf galaxies both structurally and morphologically, and that these transformations are not independent of the galaxy mass.

\subsection{What causes the cluster-centric relations?}

Unfortunately the current large-scale hydrodynamical simulations, like Illustris-TNG, cannot accurately model the galaxies with stellar masses of M < 10$^{8}$ M$_{\odot}$ due to resolution issues. However, we can still try to understand the mechanisms that are taking place in the Fornax cluster. If we make a simple assumption that the transformation of galaxies from star forming late-types to the quiescent early-types happens only through harassment or gas-stripping, what would we expect to see in the dwarf galaxy populations?

\indent In the case of harassment, the high velocity encounters strip off material from the galaxies, but it is relatively inefficient in removing  a significant number of stars (\citealp{Smith2015}). However, even if the harassment is not very important, it is known to slowly increase the S\'ersic $n$ of the galaxies, increase their surface brightness in the center, and make them redder, as the star formation stops (\citealp{Moore1998}, \citealp{Mastropietro2005}). The observational consequences of this process would then be that at a given mass, the galaxies become redder with brighter $\bar{\mu}_{e,r'}$, and increasing $C$. Harassment also heats the stars in a galaxy, resulting in thickening of its disk.

\indent In the case of ram-pressure stripping, the stars are not affected, and thus their evolution after the gas-removal happens via fading of their stellar populations. A direct observational consequence of this would be that a galaxy with a given mass should become redder with fading $\bar{\mu}_{e,r'}$, while for $C$ no correlation with color is expected. 

\indent A caveat in distinguishing these two scenarios from each other is that in a case where ram-pressure stripping is unable to remove all the gas from a galaxy, it will start working from outside in. This means that the gas is first stripped from the outer parts of the galaxy, and the star formation may still continue in the center. This process is able to accumulate more stellar mass into the galaxy center after its outer parts are quenched, and thus lead to higher surface brightness and increased peakedness of the luminosity profile. The stellar populations will contain also younger stars in the galaxy center with respect to the outskirts. Some dwarf galaxies in the clusters indeed have negative color gradients \citep{Urich2017}, which may indicate this process being important as well. As these two processes, harassment and partial ram-pressure stripping, leave similar imprints to the galaxies and both work most effectively in the center of the cluster, it is not possible to distinguish between them from our observational data, but we can still distinguish between complete ram pressure stripping and the other scenarios.

\indent Using our sample, we can test these theoretical predictions in a straightforward manner. In Fig. \ref{fig:sfb_vs_par}, we divide the galaxies in four luminosity bins to reduce the degeneracy caused by the luminosity correlations of the parameters, and, using these bins, compare the galaxy colors with their $C$ and $\bar{\mu}_{e,r'}$. We test the significance of the correlations between those parameters also using Spearman's rank correlation test, and show the correlation coefficients of each bin in the corresponding panels. We find that the galaxies in the highest luminosity bin (M$_{r'}$ < -16.5 mag) become redder with increasing surface brightness, whereas in the case of the less luminous galaxies, their $\bar{\mu}_{e,r'}$ fades while they get redder. The increase in $C$ and surface brightness in the high-luminosity bin with increasing g'-r' is consistent with the harassment and imperfect ram-pressure stripping scenarios, whereas the low-luminosity galaxies are in agreement with the complete ram-pressure stripping showing no clear correlation between $C$ and colors.

\begin{figure*}[ht]
    \centering
        \includegraphics[width=17cm]{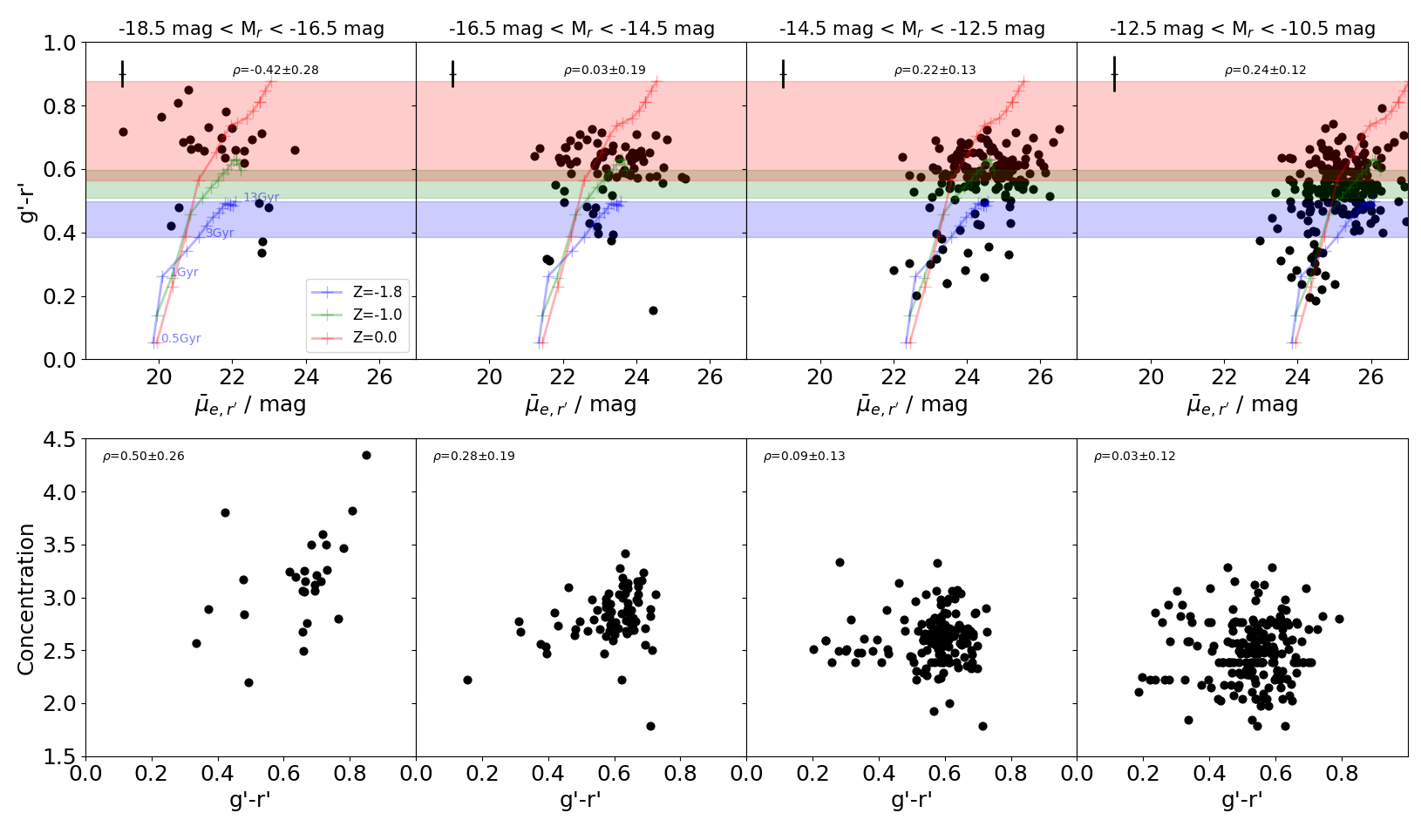}
    \caption{Upper and lower panels show the g'-r' color and the concentration parameter $C$ of the galaxies as a function of their mean efffective surface brightness in the r'-band ($\bar{\mu}_{e,r'}$), respectively. The different columns show the relations for different luminosity ranges. The lines show the evolutionary track of single stellar population for a Kroupa-like IMF with metallicities of log$_{10}$(Z/Z${\odot}$)=0.0, $-1.0$ and $-1.8$ from 0.5 Gyr to 13 Gyr (the colored dots on the lines correspond to 0.5, 1, 2, 3, 4,...,13 Gyrs). We have also shaded the color ranges corresponding the stellar population ages between 3 to 13 Gyrs with the colors matching the corresponding evolutionary tracks. The $\rho$-values reported in the upper part of the panels correspond to the Spearman's correlation coefficient and it's uncertainty calculated for the point in the corresponding panel.}
    \label{fig:sfb_vs_par}
\end{figure*}

\indent To confirm that the trend between the surface brightness and color for faint dwarfs is consistent with fading of the stellar populations, we show in Fig. \ref{fig:sfb_vs_par} the evolutionary models of a single stellar population using the models of \citet{Vazdekis2010}. We use a Kroupa initial mass function (IMF), and show the models for stellar populations with metallicities log$_{10}$(Z/Z${\odot}$)=0.0, $-1.0$ and $-1.8$ dex. The initial surface brightness of the stellar population models is set manually, but the colors are predicted purely by the model. We find that the relation between colors and $\bar{\mu}_{e,r'}$ of the M$_{r'}$ > -16.5 mag galaxies is roughly similar to the stellar population isochrones.

\indent Since the use of luminosity bins may involve some bias caused by some red and blue galaxies moving from one bin to another, we also repeat the analysis in mass bins. In Fig. \ref{fig:mu_col} and Fig. \ref{fig:mu_C}, we show that the same trends between $\bar{\mu}_{e,r'}$, g'-r', and $C$ as seen in the luminosity bins, are visible also when using mass bins. 

\begin{figure}
	\resizebox{\hsize}{!}{\includegraphics{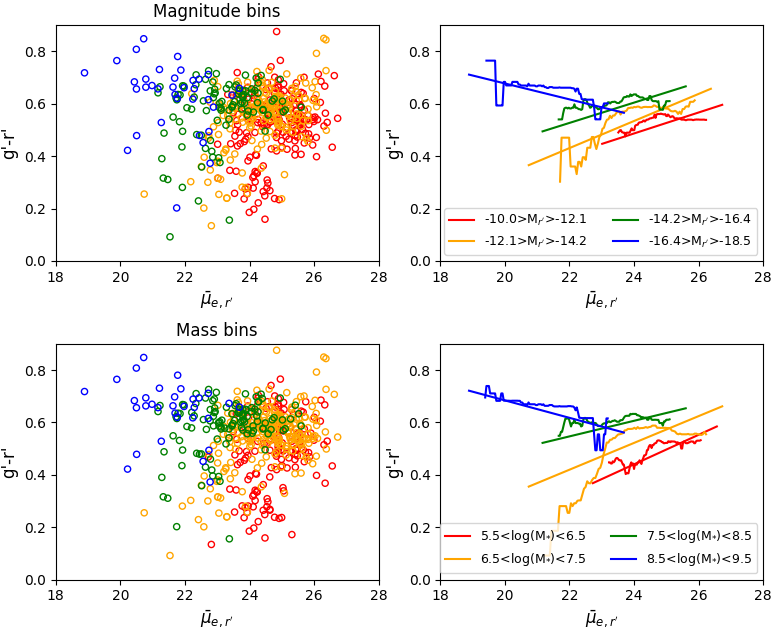}}
    \caption{Left panels show the $\bar{\mu}_{e,r'}$ versus g'-r' relation for the individual galaxies, and the right panels show the running medians and linear fits for the different bins. In the upper panels luminosity bins are used, and in the lower panels mass bins are used.}
   	\label{fig:mu_col}
\end{figure}

\begin{figure}
	\resizebox{\hsize}{!}{\includegraphics{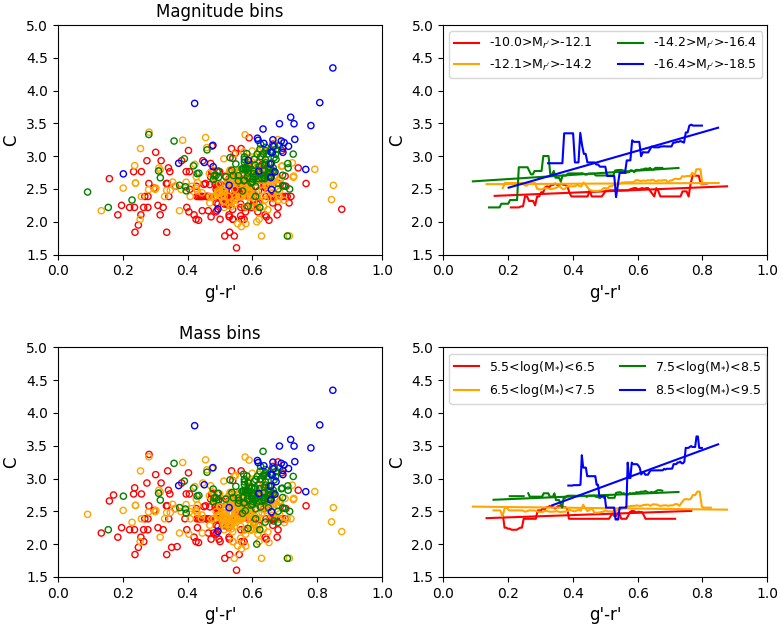}}
    \caption{Same as Fig. \ref{fig:mu_col}, but shows $C$ versus g'-r'. }
   	\label{fig:mu_C}
\end{figure}

\indent As a summary, we find that the colors, $C$ and $\bar{\mu}_{e,r'}$ of our galaxies are consistent with the scenario in which the most luminous dwarf galaxies (M$_{r'}$ < -16 mag) get quenched via harassment and/or by impartial gas stripping, since they become redder and their $C$ increases with increasing surface brightness. The quenching of the low-mass galaxies is rather consistent with complete gas-stripping and subsequent fading.  These two different mechanisms taking place in the different luminosity-ranges, can explain also why at the low-mass end (M$_{r'}$ > -14 mag) the late-type galaxies are smaller than the early-type galaxies, while in the low-luminosity end the opposite is true (see Fig. \ref{fig:mag-par-relations}).

\subsection{Effects of environmental processes}

The properties of dwarf galaxies in our sample are consistent with a scenario in which the massive dwarfs are transformed by harassment and imperfect ram-pressure stripping, and the less massive by complete ram-pressure stripping. In order to confirm that this is indeed a physically feasible scenario, we next estimate the physical strengths of the different environmental processes using analytic formulae. 

\indent In the following calculations we assume the cluster to be a static entity whose potential and galaxy density does not change. This is clearly not true in time scales of several Gyrs. However, the assumption of static cluster is necessary, since we lack information about the past evolution of the Fornax cluster and the analytical calculations would become difficult without such assumption. Thus, our following calculations are more accurate for processes that are effective in short time scales, and the results regarding longer time scales should be interpreted with caution.

\subsubsection{Ram-pressure stripping}

For estimating the strength of ram-pressure stripping, we follow the approach of \citet{Jaffe2018}. The pressure between the hot intra-cluster gas and cold gas of an infalling galaxy, can be calculated using the relative speed of the galaxy with respect to the cluster $\Delta v_{cl}$ and the density of the intra-cluster gas $\rho_{ICM}$ \citep{GunnGott1972}
\begin{equation}
P_{ram}\, = \, \rho_{ICM} (\Delta v_{cl})^2.
\end{equation} 
We estimate the radial gas density profile using the integrated gas mass profile from \citet{Paolillo2002} (Fig.17 of Paolillo et al.), which we approximate within one degree from the cluster center with a power law $M(<R) = 5\times10^5 R^{1.5}$ M$_{\odot}$, where R is in arcsec. 

\indent In order to calculate how much gas of the galaxy gets stripped, we need to calculate the anchoring force per unit area of the gas in a galaxy, $\Pi_{gal}$. The upper limit of the anchoring force at the galacto-centric distance $r'$ can be estimated by \citep{GunnGott1972} 
\begin{equation}
\Pi_{gal} (r') \, = \, 2 \pi G \Sigma_s(r') \Sigma_g(r'), 
\end{equation}
where $\Sigma_g$ and $\Sigma_s$ are the surface density profiles of the gas and stars of the galaxy, respectively. We acknowledge that this analytical form gives only an upper limit for the anchoring force and it may thus lead to overestimating the radius where the gas can be bound within the galaxy. However, for our purpose it is not important to find the exact maximum radius where the gas can be bound, but to find the mass range where the galaxies can retain gas bound at least in their central parts. For simplicity, we estimate both the gas and stellar profiles of the galaxies with an exponential disk\footnote{Exponential disk corresponds to a S\'ersic profile with $n$=1.}
\begin{equation}
\Sigma_{s/g}(r') \, = \, \Sigma_{s/g,0} \exp\left(-r'/R_d\right),
\end{equation}
where $R_d$ is the disk scale length (corresponding to 1.678R$_e$), and $\Sigma_0=M_d/(2\pi R_d^2)$, where $M_d$ is the mass of the disk. For the stellar disk $M_{d,s}$ we use the masses from Section 2.2.1 and take  R$_{e,s}$ from the r'-band S\'ersic profiles. For the gas component we use the values from \citet{Popping2014}, who found that the H$_I$ and H$_{II}$ masses for log(M$_*$/M$_{\odot}$) $\approx$ 7 galaxies are both on the order of 150\% of the stellar mass. We realize that the relative fractions of the H$_I$ and H$_{II}$ masses and their total mass both depend on the stellar-mass of the galaxy (\citealp{Catinella2018}, \citealp{Popping2014}), but as we aim to make an order of magnitude estimate, we do not take these factors into account in our model. The gas extent is approximated to be $R_{d,g}=1.7\times R_{d,s}$, as found by \citet{Cayatte1994} for the Virgo gas-rich spirals. This results into $\Sigma_{g,0}=0.3\times \Sigma_{s,0}$. 

\indent Using the $\rho_{ICM}$ and $\Pi_{gal}$ profiles, we calculate the galacto-centric stripping radii R$_s$ (where $\Pi_{gal}=P_{ram}$), at the cluster-centric distances 0.3 deg and 1 deg for galaxies traveling with velocities of $\Delta v_{cl}$ = 370 kms$^{-1}$ and 550 kms$^{-1}$. As $\Pi_{gal}$ depends on $R_d$, we use the mean $R_d$ of a given galaxy mass for calculating R$_s$. The R$_s$ as a function of the galaxy mass are shown in the left panel of Fig. \ref{fig:process_strengths}. We find that ram-pressure strips all the gas from galaxies with M$_*$ < 10$^{7.5-8}$ M$_{\odot}$ that enter the sphere within the inner 1 deg ($\approx$0.5 R$_{vir}$) of the cluster. Ram-pressure is also capable of stripping the gas of more massive galaxies at galacto-centric distances larger than r=2-3 kpc, but a remarkable finding is that ram pressure-stripping cannot strip the gas of the M$_*$>10$^8$ M$_{\odot}$ galaxies from their inner $\approx$2 kpc areas. Thus, in order to get rid of the remaining cold gas of these massive dwarfs, either star-formation has to consume the gas, or tidal interaction play an important part in removing it.

\begin{figure*}[tp]
    \centering
        \includegraphics[width=17cm]{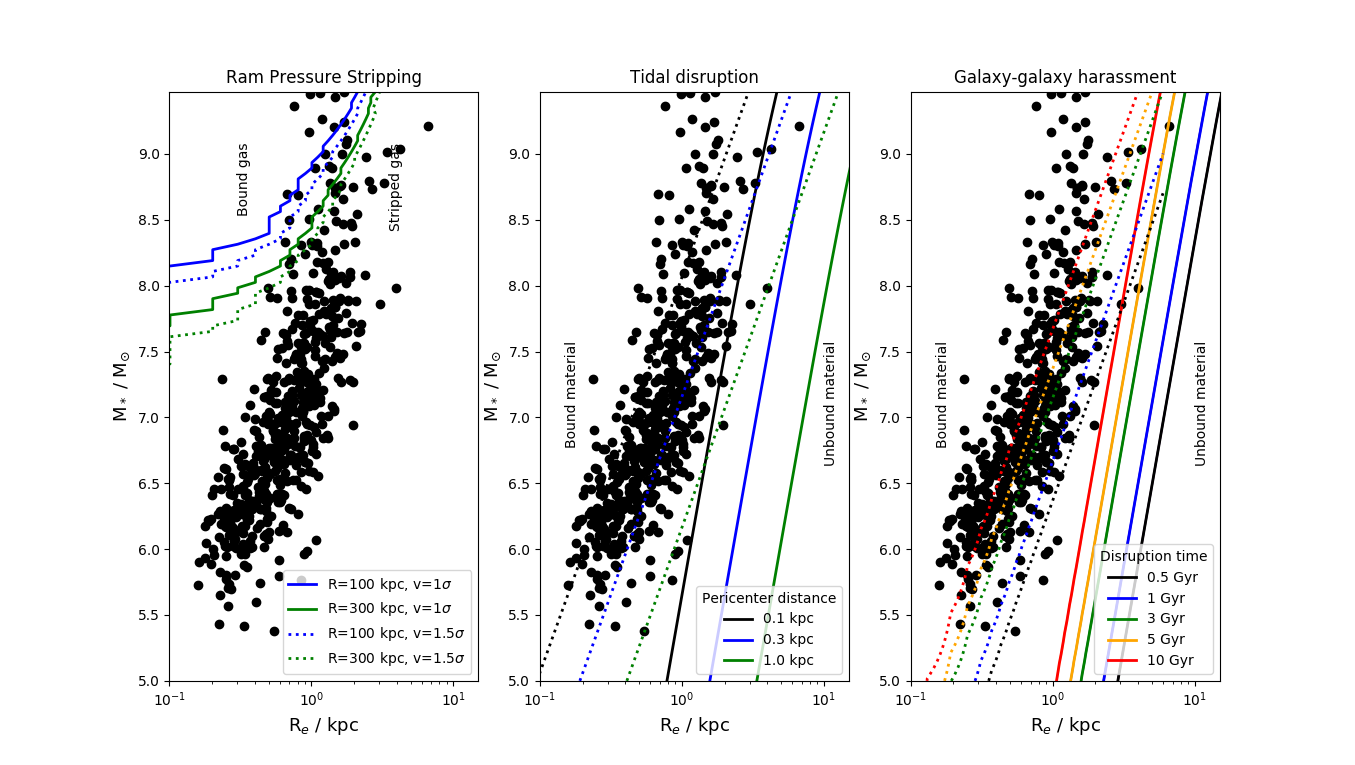}
    \caption{Effectiveness of the different environmental processes in the galaxy stellar mass-effective radius plane. The black points correspond to the FDSDC galaxies. The lines in the {\bf left panel} shows the stripping radius as a function of galaxy's stellar mass. For the areas at the right side of the lines $\Pi_{gal}<P_{ram}$, and thus all the gas gets stripped. The blue and green lines show the stripping radii at the cluster-centric distances of 100 and 300 kpc ($\approx$ 0.3 deg and 1 deg), respectively, and the solid and dotted lines correspond to galaxy velocities of 370 kms$^{-1}$ and 550 kms$^{-1}$, respectively. The lines in the {\bf middle panel} shows the the relation between the tidal radius and galaxy mass for three orbits with different peri-center radii.  The solid and dotted lines show the tidal radii for the galaxies when assuming DM-halo and without DM-halo, respectively. The solid and dotted lines in the {\bf right panel} show the disruption times against harassment for the given mass and effective radius galaxies, when assuming a dark matter halo and when using only the stellar mass for the calculation, respectively.}
    \label{fig:process_strengths}
\end{figure*}

\indent There are some caveats in the analytic ram-pressure model adopted by us (see the discussion of \citealp{Jaffe2018}). The model assumes a disk-like galaxy that is falling into the cluster face-on, and does not take into account the dark matter potential. As we showed in the Section 4.2, the late-type dwarfs have rather low intrinsic $q$ ratios, which supports the disk assumption. \citet{Roediger2006} showed that the inclination angle of the infalling galaxy does not play a major role in the significance of the gas mass-loss in the ram-pressure stripping. Finally, the mass of the dark matter halo is not as flattened as the stellar disk, and \citet{Diaz-Garcia2016} showed that only 4 \% of the total mass of the dark matter halo is within the optical part of the disk. This means that DM does not significantly alter our estimation of the anchoring force in the disk plane. However, a larger DM halo clearly makes the potential well of the galaxy deeper, and thus makes it harder for the gas to escape from the halo. We argue that even if the gas does not escape from the halo, but is pushed significantly outside the stellar disk, it is likely to be stripped due to the tidal forces in the cluster that are effective outside a few effective radii (see the next sub-sections). Thus, we can likely rely on these ram-pressure stripping approximations at least as order of magnitude estimations of the removal of cold gas from the disk plane.

\subsubsection{Tidal disruption from the cluster's tidal field}

The effects of the tidal interactions on galaxy structure are difficult to quantify without using full galaxy simulations. However, we can calculate for a given orbit and galaxy mass what is the maximum galacto-centric radius where it can keep material gravitationally bound. 

\indent The tidal radius $R_{tidal}$ is the maximum galacto-centric radius, where material can still be gravitationally bound to a galaxy. For an orbit with a cluster pericenter distance $R_{peri}$ and eccentricity $e$, the tidal radius can be calculated as follows (\citealp{King1962}, \citealp{Wittmann2016}):
\begin{equation}
R_{tidal} \, = \, R_{peri} \left(\frac{M_{obj}}{M_{cl}(<R_{peri})(3+e)} \right)^{\frac{1}{3}}, 
\end{equation}
where $M_{obj}$ is the mass of the galaxy and $M_{cl}(<R_{peri})$ is the cluster's mass within the pericenter distance of the orbit from the cluster center. For the eccentricity of the orbit, $e$, we adopt a value of 0.5\footnote{This value is not based on any observational evidence of the typical orbits in the Fornax cluster, but is rather adopted to represent an orbit between a radial and circular orbit. The results are not very sensitive for this parameter, since its variation between 0 and 1 only results into a 15\% difference between the estimated tidal radii.}.  The $R_{tidal}$ as a function of galaxy stellar mass are shown with the dotted lines in the middle panel of Fig. \ref{fig:process_strengths} for three orbits with different $R_{peri}$. It is evident that, if we only take into account the stellar mass of the galaxies, many galaxies should be tidally disrupted even on orbits with large pericenter distances. However, it is generally believed that dwarf galaxies are typically dark matter dominated, which has to be taken into account in the calculation. We adopt the stellar-to-halo mass relation of \citet{Moster2010} (see Appendix D for details) to estimate the halo masses of the sample galaxies. Since the DM-halo is more extended than the stellar component of the galaxies, we only take into account the DM-halo mass within 2R$_e$. \citet{Diaz-Garcia2016} found that by multiplying the halo-to-stellar mass relation of Moster et al. by a factor of 0.04, the relation becomes consistent with the dark matter masses within the optical radii of the S$^4$G \citep{Sheth2010} galaxies, estimated using the H$_I$ rotation amplitudes and stellar gravity fields calculated from 3.6 $\mu$m images. We thus use the Moster et al. DM masses multiplied by the factor of 0.04 in our calculations. We acknowledge that the harassment in the cluster affects also the DM halo, by stripping material out from it (\citealp{Rys2014},  \citealp{Smith2015}). Therefore our estimations for the  radii where the material can be gravitationally bound to the galaxies should be considered as upper limits. We show $R_{tidal}$ where the dark matter contribution has been taken into account shown with the solid lines in Fig. \ref{fig:process_strengths}. In this case, the material is bound to the galaxies up to the radius of r=2-3 R$_e$'s even on the orbit with the smallest pericenter distance. Analogously, this same test could be done to probe whether the dwarfs could survive in the cluster when Modified Newtonian Dynamics  is applied. However, this is out of scope of this paper.

\subsubsection{Galaxy-galaxy harassment}

Another process that disrupts galaxies in the cluster environment is harassment. Since the relative speeds of the galaxies are large ($\sqrt{2}\times370$kms$^{-1}$ $\approx$500 kms$^{-1}$) in the cluster, we can use the high-velocity impulse approximation. If the relative velocities are high enough, the potentials of the galaxies do not change during the interactions, and the internal kinetic energy of a galaxy can be assumed to change instantaneously. \citet{BinneyTremaine2008} approximate that in a galaxy cluster with the velocity dispersion of $\sigma$ and galaxy number density of $n_p$ of $M_p$ mass galaxies, the disruption time for galacto-centric radius $a$, can be calculated as (see their Eq. 8.54)
\begin{equation}
t_d \, = \, \frac{0.043}{W}\frac{\sqrt{2}\sigma M_s r_h^2}{G M_p^2 n_p a^3},
\end{equation}
where $M_s$ is the mass of the subject galaxy, $r_h$ is the perturbers' half mass radius, and $W$ is a parameter related to the shape of the perturbing galaxy's mass distribution (we assume $W=1$). We obtain $n_p$=25 Mpc$^{-3}$ by calculating the number density of the giant galaxies (M$_{r'}$ < -18.5 mag) within the virial radius of the cluster. We take the masses and effective radii for those galaxies from \citet{Iodice2018} whenever available, and for those with no measurements by Iodice et al., we use the magnitudes and effective radii from the FCC. From the magnitudes we estimate the stellar masses of those galaxies, and use them for obtaining the halo masses using the stellar-to-halo mass relation of Moster et al. For $M_p$, we take the median of those galaxy masses $M_p$ = $10^{11.6}$ M$_{\odot}$. Since we cannot measure $r_h$ from the data, we need to obtain the relation between $r_h$ and R$_e$ from the simulations. We select a cluster with a similar mass as the Fornax cluster has from the Illustris simulation \citep{Pillepich2018}, and take the median $r_h$/R$_e$ of the giant galaxies in that cluster. More precisely, we selected the giant galaxies of the Illustris group 11 (M$_{vir}$=7.9$\times10^{13}$ M$_{\odot}$), and found the median $r_h$/R$_e$ = 3.6. For $r_h$ we thus multiply the median R$_e$ of the Fornax giant galaxies, and obtain a value of $r_h$ $\approx$14 kpc.

\indent In the right panel of Fig. \ref{fig:process_strengths}, we show the disruption times for the given mass and R$_e$ of the galaxies. We show again the disruption times with and without the dark matter halo contributions, indicated with the solid and dotted lines, respectively. We find that a large fraction of galaxies would be short lived without any dark matter halos. While taking into account the dark matter contribution, we find that most of the galaxies can keep the stars bound within their R$_e$ for  more than 10 Gyr. Especially the low-mass dwarfs are all very resistant against the harassment due to their high M/L-ratios.

\subsubsection{Are theory and observations consistent?}

In Sections 5.3 and 7.2, we discussed that the properties of the galaxies  with M$_*$ > 10$^8$ M$_{\odot}$ are consistent with them being affected by harassment and outer disk gas stripping in the cluster environment, whereas the galaxies smaller than that are consistent with being mostly quenched by ram-pressure stripping. By calculating the strengths of the different environmental processes, we found that ram-pressure stripping is indeed efficient in removing all the gas from M$_*$ < 10$^8$ M$_{\odot}$ galaxies, but can not remove the gas in the innermost $\approx$1 kpc of the galaxies more massive than that. As ram-pressure stripping quenches the galaxy more rapidly than the other processes, our finding that the early-type fraction in the high-luminosity end of the luminosity function is lower than in the low-mass end (Fig. \ref{fig:nuc_frac}), seems to be compatible with the  scenario that the high-mass galaxies will remain longer as late-type than the low-mass galaxies. We found that tidal disruption of galaxies due to the cluster potential is very rare, but over time the galaxy-galaxy interactions cause the outskirts of the most extended galaxies to disrupt. Indications of such interactions have also been observed for the giant ETGs in the center of the cluster (\citealp{Iodice2018}). 

\indent The physical processes that we suggest to explain our observations in the Fornax are also consistent with the conclusions of the kinematic studies of dwarf galaxies in the Virgo cluster (\citealp{Rys2014}, \citealp{Toloba2015}). \citet{Toloba2015} found that the the specific stellar angular momentum dwarf galaxies in the Virgo cluster decreases toward the inner parts of the cluster and that the dEs which rotate fast have often disk structures like spiral arms. \citet{Rys2014} also found that dwarfs in the central parts of the Virgo have lower M/L-ratios than those in the outskirts. Those observations can be explained with the scenario where ram-pressure stripping quickly quenches majority of the galaxies, and harassment then slowly rips off DM from the outer parts of their halos and heats up the stellar disks of the galaxies. In the Fornax cluster similar studies have not yet been made.

\indent For the low-mass galaxies the observed properties are well in agreement with the ram pressure-stripping scenario. All the low-mass dwarfs will lose their gas when they enter the inner 1 deg of the cluster, and their star formation will thus stop and stellar populations start fading. Harassment seems to be less important for these galaxies since they reside in massive DM haloes. However, in 10 Gyrs, these galaxies start to lose their material outside the galacto-centric radii of 2 kpc, which corresponds to several effective radii. For an exponential profile the stellar mass outside that radius is small.

\indent The gas of the M$_*$ > 10$^8$ M$_{\odot}$ galaxies is also removed from their outer parts by ram-pressure stripping, but they can not be fully quenched just by that process. Thus to fully deplete the gas of a massive dwarf galaxy it must experience harassment or transform the remaining cold gas in the center into stars. This is also consistent with our observations. We acknowledge that also internal processes like feedback from active galactic nuclei and supernovae (\citealp{Silk2017}, \citealp{Dashyan2018}) can play a role in the quenching of dwarfs in the different mass ranges, but as those processes are yet poorly constrained we focus our discussion to the external processes only. As the gas depletion time scales by ram-pressure stripping are longer then the crossing time scales, it is likely that there are some galaxies with central star formation also in the cluster center. However, according to the galaxy density profiles in Fig. \ref{fig:radial_distribution} such galaxies seem to be rare. The disruption time scales also show that the outer R > 5 kpc parts of the massive dwarf galaxies will be tidally disrupted in about 5 Gyr, which also explains why we do not see many galaxies with such large $R_e$ in the Fornax cluster. To differentiate the effects between the ram-pressure stripping and harassment in massive dwarfs, it would require to study the radial chemical abundances of the stellar populations and look at the radial surface brightness profiles in detail for truncations. This is out of reach of this study, and should be studied in future. 

\indent A point that should be considered also in the case of these massive dwarfs, is that their progenitors have likely had slightly different size than the present day late-types in the Fornax cluster. \citet{VanderWel2014} show that the median R$_e$ of galaxies with stellar mass 10$^9$ M$_{\odot}$ < M$_*$ < 10$^9.5$ M$_{\odot}$ has grown by $\sim$ 30\% since z=1. This indicates that the environmental effects are only partially responsible for the contraction of R$_e$s of massive dwarfs from the outer to inner parts of the cluster, and it is partly also resulting from the fact that the progenitors of the innermost dEs were smaller than the late-types that we observe now falling into the cluster.

\indent In Section 4.2, we did not find significant changes in the luminosity function in different parts of the Fornax cluster. This finding is consistent with the theoretical expectation that not many galaxies should be disrupted due to the cluster potential, and that the disruption time scales are mostly long even within the inner parts of the Fornax cluster. 

\indent To conclude, our observations and the theoretical predictions of the efficiency of the different environmental processes are in agreement. Qualitatively, these environmental processes can explain the cluster-centric distribution of different types of dwarfs (Section 3), shapes of the parameter scaling relations and the cluster-centric trends (Section 5), as well as the changes in the galaxy colors (Section 6). Clearly, it is necessary to test these results with more elaborated simulations and follow-up observations.

\subsection{Are the dwarf galaxies in the Virgo and Fornax clusters different?}

The most complete dwarf galaxy sample in terms of luminosity  in the Virgo cluster is the one of the NGVS \citep{Ferrarese2012}. The multifilter observations of the NGVS have similar depth and resolution as ours. What makes the comparison difficult is that, up to now, the NGVS collaboration has only published the results for the core part of the cluster. 

\indent In the previous sections, we have shown that the color-magnitude relations of the early-type galaxies are similar in the cores of the Fornax and Virgo clusters, the axis-ratio distributions are similar, and that the faint end of the luminosity functions have similar shapes in these clusters. Additionally, the segregation in the spatial distributions of the galaxies with different morphological types has also been observed in both clusters. Thus, there is no observational evidence that the dwarf galaxies in these two clusters are different.

\indent As the Virgo and Fornax clusters have different total masses and Virgo seems to contain more sub-structure than Fornax, it is surprising that we did not find differences in their dwarf populations. The galaxy number density in the center of the Fornax is two times higher than in the Virgo cluster \citep{Jordan2007}, and the velocity dispersion in the Virgo cluster is two times higher $\approx$ 700 km s$^{-1}$ \citep{Binggeli1993} than in Fornax, making the harassment time scales in the Virgo cluster four times longer with respect to the Fornax cluster. The mass concentration around M87 \citep{Fabricant1980} in the center of the Virgo cluster is only slightly higher than the one around NGC 1399, which makes the tidal disruption of the dwarfs similar around these two galaxies. On the other hand, the density of the X-ray gas in the center of Virgo \citep{Fabricant1980} is roughly five times higher than in the Fornax cluster \citep{Paolillo2002} which, combined with the two times higher velocities of the Virgo dwarfs, makes ram-pressure in the Virgo cluster $\approx$20 times more efficient than in the Fornax cluster. By converting the gas-density and galaxy velocity dispersion of the Virgo cluster into the ram-pressure stripping model of Section 7.3.1, we find that in Virgo, ram-pressure stripping should be able to deplete all the gas from three times more massive galaxies than in the Fornax cluster.

\indent Since the low mass dwarf galaxies (log$_{10}$(M$_*$/M$_{\odot}$)<7.5) seem not to be effected by harassment even in the Fornax cluster due to their massive DM haloes, and they are likely quenched by ram-pressure stripping in both clusters, there should be no difference in their properties due to the environmental effects discussed here. Since ram-pressure stripping is effective for more massive galaxies in the Virgo cluster, we should see there more dEs with larger effective radii (R$_e$>2kpc) than in the Fornax cluster with masses 7.5<log(M$_*$/M$_{\odot}$)<8.5. These galaxies that fall into the category of Ultra Diffuse Galaxies (UDG), have been detected both in the Fornax (\citealp{Bothun1991}, \citealp{Munoz2015},\citep{Venhola2017}) and Virgo clusters (\citealp{Sandage1984},  \citealp{Mihos2015}), but systematic studies of the abundance of UDGs in both clusters are still lacking. In order to perform a more quantitative and less speculative comparison between these clusters we need to wait for the publication of the full sample of the NGVS.

\section{Summary and conclusions}

In this paper, we have analyzed the photometric properties of the dwarfs in the FDSDC. The used sample consists of 564 dwarf galaxies within the stellar mass range of 2$\times10^{5}$  M$_{\odot}$ < M$_{*}$ < 2$\times10^{9}$ M$_{\odot}$ corresponding to r'-band absolute magnitude range of -9 mag > M$_{r'}$ > -18.5 mag. Using the photometric measurements and morphological classifications of Venhola et al. (2018), and the statistical analysis in this work, we wanted to understand how the properties of the dwarf galaxies are affected by the cluster environment, and what are the most important processes transforming the galaxies.

\indent We showed that the cluster environment changes the morphology of the galaxies from late-type to early-type, manifested as an increasing early-type fraction and a drop of late-types toward the inner parts of the cluster. We also showed evidence that the environmental transformation happens via different mechanisms depending on the mass of the galaxy. The most massive dwarf galaxies (M$_{r'}$ < -16 mag) became more concentrated and redder towards the center of the cluster, which is consistent with the effects of harassment and outer disk gas stripping. The less luminous galaxies are redder in the cluster center but they have systematically lower surface brightnesses than the population in the outer parts of the cluster. These trends are consistent with the scenario in which star formation is shut down via complete gas-stripping. The further evolution is explained by a passive fading of the stellar populations. We also showed that the theoretical predictions of the effects of harassment and ram-pressure stripping in the Fornax cluster, are consistent with our observations, thus giving further support for the theory.

Our main conclusions are:

\begin{itemize}
\item We find that morphology and spatial distribution of the dwarf galaxies in the Fornax cluster are strongly correlated, so that the most concentrated are dE(N)s, then dE(nN)s, and least concentrated are the late-type dwarfs (Fig. \ref{fig:radial_distribution}).  By deprojecting the radial galaxy surface number density profiles into 3D number density profiles, we have shown that the cluster-center is devoid of late-type dwarfs and dE(nN)s. 
\vskip 0.5cm

\item The early-type fraction and nucleation fraction increase towards the center of the main cluster (Fig. \ref{fig:type_fractions}). In Fornax A the fractions of nucleated and early-type galaxies are lower than in the main cluster and they do not show clear correlation with the group-centric distance.
\vskip 0.5cm

\item The late-type dwarfs and early-type dwarfs have similar luminosity functions, but only if both dE(N)s and dE(nN)s are considered into the early-type population (Fig. \ref{fig:morphology_luminosity_func}). The nucleation fraction is a strong function of magnitude and it reaches its peak around M$_{r'}$ = -16 mag, and then quickly declines towards the low luminosity end (Fig. \ref{fig:nuc_frac}). 
\vskip 0.5cm

\item The luminosity function of the dwarf galaxies does not significantly change in the different parts of the cluster (Fig. \ref{fig:luminosity_function}). We also showed that this is not in contradiction with the theoretical predictions, since the cluster tidal field in the center of the Fornax cluster can only disrupt galaxies on orbits with small cluster-centric pericenters. The luminosity function in the Fornax cluster is also identical to those in the Virgo and Coma clusters, but has steeper low mass end than the luminosity function in the Local Group, if the completeness correction is applied. 
\vskip 0.5cm

\item We find that the size (R$_e$), colors, peakedness (S\'ersic $n$), and surface brightness ($\bar{\mu}_{e,r'}$) of the early- and late-type galaxies in the Fornax cluster follow different relations with respect to the absolute magnitude of the galaxies (Fig. \ref{fig:mag-par-relations}). Their shape distribution differs as well (Fig. \ref{fig:ba_analysis}). However, dE(N)s have identical colors and sizes as the dE(nN)s of a given mass. 
\vskip 0.5cm

\item We show that the color-magnitude relation of the dEs in our sample does not show statistically significant variations with respect to the cluster-centric distance in the colors that do not include u'-band (Fig. \ref{fig:cm_relations_var}). When considering the colors including the u'-band, we find that the dEs become significantly redder towards the inner parts of the cluster, indicating that the galaxies in the outer parts were quenched more recently that those in the inner parts.
\vskip 0.5cm

\item We find that in luminosity bins the low-luminosity galaxies (M$_{r'}$ > -16 mag) become redder, more extended and have lower surface brightnesses towards the center of the cluster (Fig. \ref{fig:distance_vs_magnitude}). The most luminous dwarf galaxies (M$_{r'}$ < -16 mag) become more peaked and redder towards the inner parts of the cluster. The result that the low-luminosity dwarf galaxies become redder with decreasing surface brightness seems to be consistent with an evolution where the gas reservoirs of these galaxies are fully removed by gas stripping, and they then simply fade to red and old systems. The galaxies with M$_{r'}$ < -16 mag become redder with increasing surface brightness, which is consistent with both harassment and imperfect ram-pressure stripping, in which their inner parts continue to form stars longer than their outskirts. It is likely that both of these processes take place. These observations are in good agreement with our calculations in Section 6.2, which showed that the ram-pressure stripping can not remove all the gas from the most massive dwarfs and thus they need to be quenched some other way, possibly by tidal interactions or by feedback from active galactic nuclei.
\vskip 0.5cm

\end{itemize}

{\bf Acknowledgements:} We would like to thank Thorsten Lisker for the valuable help and discussions during the whole FDS project. A.V. would like to thank the Vilho, Yrjö, and Kalle Väisälä Foundation of the Finnish Academy of Science and Letters for the financial support during the writing of this paper. GvdV acknowledges funding from the European Research Council (ERC) under the European Union’s Horizon 2020 research and innovation program under grant agreement No 724857 (Consolidator Grant ArcheoDyn). R.F.P., T.L., E.L., H.S., E.I., and J.J. acknowledge financial support from the European Union’s Horizon 2020 research and innovation program under the Marie Skłodowska-Curie grant agreement No. 721463 to the SUNDIAL ITN network. H.S., E.L., and A.V. are also supported by the Academy of Finland grant n:o 297738. C.W. is supported by the Deutsche Forschungsgemeinschaft (DFG, German Research Foundation) through project 394551440. J.F-B acknowledges support from grant AYA2016-77237-C3-1-P from the Spanish Ministry of Economy and Competitiveness (MINECO).
\bibliographystyle{aa}

\bibliography{dwarf_paper}

\begin{appendix}
\section{Completeness of the FDSDC}

Since the FDSDC is a size and magnitude limited galaxy sample we show in Fig. \ref{fig:detection_completeness} how these selection limits restrict our analysis to a certain area in the size-luminosity parameter space. The galaxies are detected using SExtractor, which uses minimum isophotal size threshold for detecting objects. \citet{Ferguson1988} showed that if a sample is limited by a minimum diameter $d_{lim}$ at isophote $\mu_{lim}$, a galaxy with an exponential light profile and effective radius R$_e$ should have a total apparent magnitude  
\begin{equation}
m_{tot} < \mu_{lim} - \frac{r_{lim}}{0.5487\mathrm{R}_e}-2.5\log_{10}\left[ 2\pi\left(0.5958\mathrm{R}_e\right)^2\right],
\end{equation}
in order to be selected into the sample. For the FDSDC $d_{lim}$ = 4 arcsec and $\mu_{lim}$ $\approx$ 26.5 mag arcsec$^{-2}$. In Fig. \ref{fig:detection_completeness} we show this relation with the solid line. 

\begin{figure}
	\resizebox{\hsize}{!}{\includegraphics{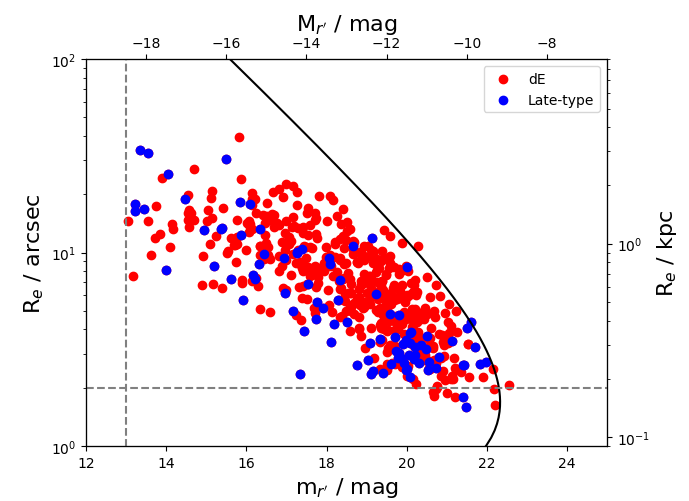}}
    \caption{ Total r'-band apparent magnitudes (m$_{r'}$) and effective radii (R$_e$) are shown for the late-type (blue symbols) and early-type (red symbols) galaxies of the FDSDC. The vertical and horizontal dashed lines show the minimum magnitude (M$_{r'}$ > -19 mag) and minimum size limits ($a$ > 2 arcsec) of the sample, respectively. The solid line shows the selection curve resulting from the SExtractor detection. An object can be detected is it has diameter larger than 4 arcsec at the 26.5 mag arcsec$^{-2}$ isophote. The upper x-axis and the right y-axis show the absolute magnitudes and effective radii of the galaxies at the distance of the Fornax cluster.}
   	\label{fig:detection_completeness}
\end{figure}

\section{Onion peeling deprojection method}

In order to analyze 3D radial distributions of galaxy number densities instead of projected distributions, we derived deprojected distributions from the observed ones, by using the "onion peeling" method \citep{Fabian1981,Kriss1983}. In order to perform this transform, we need to do two assumptions:

\begin{itemize}
\item {\bf Spherical symmetry : } We assume that the galaxies are distributed with spherical symmetry around the cluster.

\item {\bf No blending bias : } We assume that we see all the galaxies within our selection limits, so that we do not need to estimate how many galaxies are hiding behind large galaxies.

\end{itemize}

Both of the above mentioned assumptions can be argued against, but they are necessary to do in order to perform the deprojection. However, since Fornax (especially within the virial radius) is fairly compact and relaxed system the assumption of spherical symmetry is not completely unrealistic. Also, since the galaxy density in the central parts is $\approx$ 50 galaxies deg$^{-2}$ and a typical size for a galaxy is R$_e$ $\approx$ 10 arcsec that corresponds to area of 2.5$\times$10$^{-5}$ deg$^2$, blending seems to not be a severe problem in our case.

\indent We first assume that the cluster can be described as a set of concentric spherical layers with constant densities $\rho_i$. We can only observe the projected densities $\theta_i$ at the different cluster-centric radii, but using the spherical geometry of our cluster model we can deduce the intrinsic structure. If our bins have radial width of $h$, we can write the projected density of a bin $k$ as:
\begin{equation}
\theta_k = \frac{ \Sigma_{i,i \geq k} \, V_i^{k} \, \rho_i }{\pi \left[ (r_k+\frac{h}{2})^2 - (r_k-\frac{h}{2})^2 \right]},
\end{equation}
where $r_k$ is the radius of the bin $k$, and $V_i^{k}$ is the volume of the intersection of a sphere with radius $r_i$ and a co-centric hollow cylinder with edge thickness of $h$ and outer radius of $r_k+h/2$. The density of the outermost bin i=$o$ can be straightforwardly obtained from the observations as:
\begin{equation}
\theta_o = \frac{ V_o^{o} \, \rho_o }{\pi \left[ (r_o+\frac{h}{2})^2 - (r_o-\frac{h}{2})^2 \right]} \, \rightarrow \, \rho_o = \frac{\theta_k \pi \left[ (r_o+\frac{h}{2})^2 - (r_o-\frac{h}{2})^2 \right]}{V_o^{o}}
\end{equation}
and it can be used to solve the other densities iteratively from outwards-in.

\indent For estimating the uncertainty, we did 100 Monte Carlo simulations of the observed radial distribution, and in each realization we varied the number of galaxies in the bin assuming a Poisson uncertainty. We ran the onion peeling deprojection methods for each Monte Carlo realization, and estimated the uncertainty by taking the standard deviation of the results. 

\indent To confirm that the method works, we tested it with a Plummer sphere \citep{Plummer1911}, for which the volume density is related to radius $r$ as follows (we used $M=a=1$):
\begin{equation}
\rho(r) \, = \, \left( \frac{3 M}{4\pi a^3} \right)\left(1+\frac{r^2}{a^2} \right)^{-\frac{5}{2}},
\end{equation}
and the projected surface density can be solved \citep{Dejonghe1987}:
\begin{equation}
\rho_p(r_p) \, = \, \frac{M}{\pi a^2(1+\frac{r_p^2}{a^2})^2}.
\end{equation}
We applied the onion peeling method for the analytical surface density profile, and solved the corresponding volume density. The results are shown in Fig. \ref{fig:plummer_test}. It is clear that the method is capable of solving the volume density very accurately, only differing slightly in the very outer parts. This small deviation is expected, since the analytical model extends to infinity, whereas the onion peeling method has to be started from some finite radius.

\begin{figure}
	\resizebox{\hsize}{!}{\includegraphics{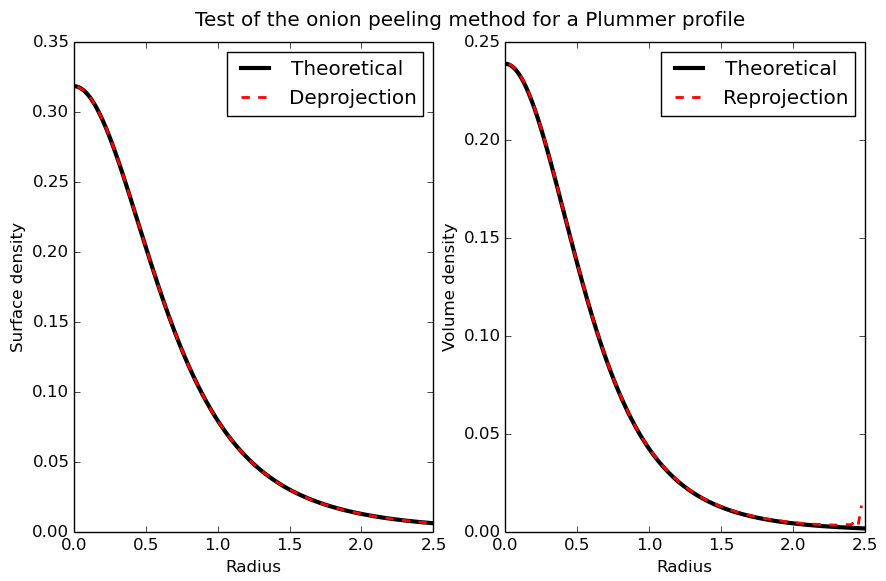}}
    \caption{Test of the onion peeling method for a Plummer sphere. The black line in the left panel shows the analytically calculated surface density for a Plummer sphere as a function of radius. The black line in the right panel show the corresponding analytical volume density. The red dashed lines in the right and left panels show the deprojected volume and surface density, respectively, obtained from the analytical surface density profile using the onion peeling method. }
   	\label{fig:plummer_test}
\end{figure}

\section{Method for obtaining intrinsic axis ratio distributions}

In Section 4, we analyze the observed axis-ratio ($b/a$) distribution, and use it to obtain the intrinsic axis ratios. We used the analytic approach of \citet{FallFrenk1983} (see also \citealp{Lisker2007}) to obtain the intrinsic distributions. In order to use the analytic formulae, we need to assume that all the galaxies have either oblate or prolate spheroidal shape. Assuming oblate galaxies with observed $b/a$-distribution function $\phi(p)$, where $p$ is the observed $b/a$, we can obtain the intrinsic axis ratio distribution function as follows:
\begin{equation}
\psi(q)\, =\, \frac{2}{\pi\sqrt{1-q^2}} \frac{\mathrm{d}}{\mathrm{d}q} \int_0^q \mathrm{d}p \frac{\phi(p)}{\sqrt{p^2-q^2}},
\end{equation}
where $q$ is the intrinsic axis ratio. When prolate intrinsic shapes are assumed, the intrinsic distribution can be obtained as
\begin{equation}
\psi(q)\, =\, \frac{2}{\pi q^2 \sqrt{1-q^2}} \frac{\mathrm{d}}{\mathrm{d}q} \int_0^q \mathrm{d}p \frac{p^3\phi(p)}{\sqrt{p^2-q^2}}.
\end{equation}
\indent In order to avoid the singularity in the formulae, we apply a variable change $x=p/q$. This variable change allows us to write the integral in form 
 $\int \mathrm{dx} f(x) (1-x^2)^{-1/2}$. Using  Gauss-Chebyshev quadrature for the integral, we can write in the form:
\begin{equation}
\int_0^1 \mathrm{dx} \frac{f(x)}{\sqrt{1^2-x^2}}= \Sigma_{i \in A}w_i f(x_i), \, \mathrm{where} \, A=\left\{ i|x_i>0 \right\}, 
\end{equation}
with
\begin{equation}
x_i=\cos\left(\frac{2i-1}{2n}\pi\right),
\end{equation}
and
\begin{equation}
w_i=\frac{\pi}{n}
\end{equation}
We used $n=100$, for calculating the integrals. When deriving $\phi(p)$, from the observed $b/a$ distribution, we smooth the distribution using convolution. In the convolution we use a Gaussian filter with $\sigma$=0.075. 

\section{Halo-to-stellar mass relation}

In Section 7, we use the stellar-to-halo mass relation from \citet{Moster2010}. The used relation between the stellar mass of a galaxy $M_*$ and its dark matter halo mass $M_{halo}$ is
\begin{equation}
\frac{M_*(M_{halo})}{M_{halo}} \, = \, 2 \left( \frac{M_*}{M_{halo}}\right)_0\left[ \left( \frac{M_{halo}}{M_I} \right)^{-\beta} + \left( \frac{M_{halo}}{M_I} \right)^{\gamma} \right]^{-1},
\end{equation}
where $\left(\frac{M_*}{M_{halo}}\right)_0$ is a normalization factor, $M_I$ is a characteristic mass, and $\beta$ and $\gamma$ are the slopes in the low and high mass ends, respectively. The constants in the relation were obtained by Moster et al. by applying abundance matching between the theoretical halo-mass function and observed galaxy-mass function and then fitting the stellar-to-halo mass function. The obtained constants were $\log_{10} \left(\frac{M_I}{M_{\odot}}\right)$=11.884, $\left(\frac{M_*}{M_{halo}}\right)_0$ = 0.02820, $\beta$ = 1.057, and $\gamma$ = 0.556. In Fig \ref{fig:halo-to-stellarmass}, we show the halo-to-stellar mass relation used in this work.

\begin{figure}
	\resizebox{\hsize}{!}{\includegraphics{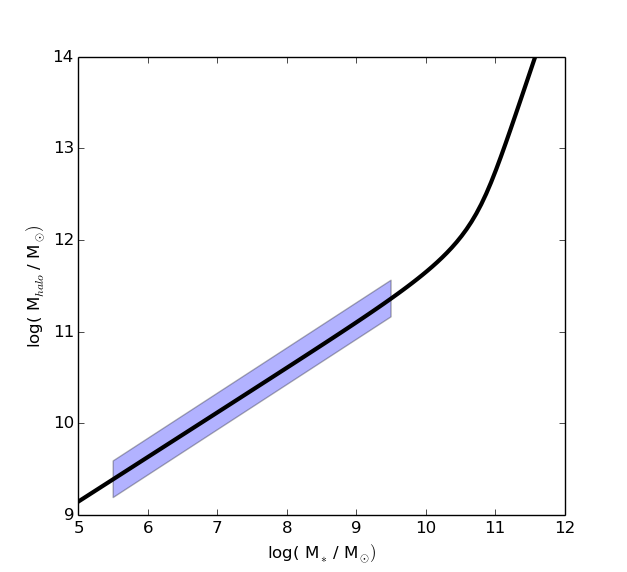}}
    \caption{Relation between the stellar mass of a galaxy and its dark matter halo mass. The black line shows the Eq. D.1 \citep{Moster2010}, and the blue shaded area indicates the mass range of the FDSDC galaxies.}
   	\label{fig:halo-to-stellarmass}
\end{figure}

\section{Transformations between photometric filters}

In Section 2.3 we transformed the Johnson R-band magnitudes and V-I filter colors into the SDSS g', r', and i' bands using the transformations from \citet{Jordi2006}. The used transformation between the R-band magnitude, $M_R$, and r'-band magnitude, $M_{r'}$, is:
\begin{equation}
M_R-M_{r'}   =  -0.153 \times (r'-i') - 0.117,
\end{equation}
where $r'-i'$ corresponds to the SDSS color. The V-I color, $V-I$, was transformed to $g'-i'$ as follows:
\begin{equation}
V-I \,  = \, 0.675\times(g'-i')  + 0.364,.
\end{equation}

\section{Table of values in Fig. 15}

\onecolumn
\begin{landscape}
\centering
\begin{table*}
\caption{Mean values of the bins in Fig. \ref{fig:distance_vs_magnitude} and the corresponding uncertainties. The rows of the table show the means of g'-r' colors, effective radii (R$_e$), S\'ersic indices ($n$), residual flux fractions ($RFF$), and mean effective surface brightnesses ($\bar{\mu}_{em.r'}$) in the magnitude and spatial bins. The mean values are followed by their uncertainties. We show the bin means calculated from the raw data (R), and from the sample of galaxies remaining after the projection bias (see Section 5.1) is taken in account (B). The columns separated with vertical lines correspond to the magnitude bins, and the sub-columns referred as 'in', 'mid', and 'out', correspond to the spatial bins defined in the Section 4.2: R < 1.5 deg from NGC 1399 (in), 1.5 deg < R < 2.5 deg (mid), and R > 2.5 deg (out).
}
\label{table:tab_fig15}
\centering
\resizebox{24cm}{!}{
\begin{tabular}{ l | c c c | c c c | c c c | c c c |} 
& \multicolumn{3}{c|}{-18.5 mag < $M_{r'}$ < -16.5 mag} & \multicolumn{3}{c|}{-16.5 mag < M$_{r'}$ < -14.5 mag}  & \multicolumn{3}{c|}{-14.5 mag < M$_{r'}$ < -12.5 mag}  & \multicolumn{3}{c|}{-12.5 mag < M$_{r'}$ < -10.5 mag} \\ \hline 
 & in & mid & out & in & mid & out & in & mid & out & in & mid & out \\ \hline 
g'-r' (B) & 0.72 $\pm$ 0.03 & 0.66 $\pm$ 0.02 & 0.57 $\pm$ 0.04 & 0.63 $\pm$ 0.009 & 0.57 $\pm$ 0.02 & 0.44 $\pm$ 0.06 & 0.59 $\pm$ 0.009 & 0.55 $\pm$ 0.02 & 0.55 $\pm$ 0.02 & 0.56 $\pm$ 0.02 & 0.52 $\pm$ 0.01 & 0.49 $\pm$ 0.02  \\  
g'-r' (R) & 0.61 $\pm$ 0.05 & 0.66 $\pm$ 0.02 & 0.57 $\pm$ 0.04 & 0.6 $\pm$ 0.01 & 0.57 $\pm$ 0.02 & 0.44 $\pm$ 0.06 & 0.57 $\pm$ 0.01 & 0.55 $\pm$ 0.02 & 0.55 $\pm$ 0.02 & 0.53 $\pm$ 0.01 & 0.52 $\pm$ 0.01 & 0.49 $\pm$ 0.02  \\ \hline 
R$_e$ (B) & 11.5 $\pm$ 1.8 & 15.1 $\pm$ 1.5 & 20.9 $\pm$ 2.8 & 13.5 $\pm$ 0.9 & 12.9 $\pm$ 0.9 & 11.3 $\pm$ 1.3 & 9.8 $\pm$ 0.9 & 9.3 $\pm$ 0.5 & 8.7 $\pm$ 0.7 & 7.0 $\pm$ 0.4 & 4.8 $\pm$ 0.3 & 4.7 $\pm$ 0.3  \\  
R$_e$ (R) & 18.1 $\pm$ 3.0 & 15.1 $\pm$ 1.5 & 20.9 $\pm$ 2.8 & 13.6 $\pm$ 0.9 & 12.9 $\pm$ 0.9 & 11.3 $\pm$ 1.3 & 9.1 $\pm$ 0.5 & 9.3 $\pm$ 0.5 & 8.7 $\pm$ 0.7 & 5.6 $\pm$ 0.2 & 4.8 $\pm$ 0.3 & 4.7 $\pm$ 0.3  \\ \hline 
$n$ (B) & 1.8 $\pm$ 0.2 & 1.6 $\pm$ 0.1 & 0.96 $\pm$ 0.1 & 1.2 $\pm$ 0.06 & 1.1 $\pm$ 0.05 & 1.1 $\pm$ 0.1 & 0.92 $\pm$ 0.05 & 0.84 $\pm$ 0.02 & 0.86 $\pm$ 0.04 & 0.94 $\pm$ 0.1 & 0.82 $\pm$ 0.03 & 0.79 $\pm$ 0.04  \\  
$n$ (R) & 1.5 $\pm$ 0.1 & 1.6 $\pm$ 0.1 & 0.96 $\pm$ 0.1 & 1.1 $\pm$ 0.04 & 1.1 $\pm$ 0.05 & 1.1 $\pm$ 0.1 & 0.88 $\pm$ 0.02 & 0.84 $\pm$ 0.02 & 0.86 $\pm$ 0.04 & 0.81 $\pm$ 0.03 & 0.82 $\pm$ 0.03 & 0.79 $\pm$ 0.04  \\ \hline 
$RFF$ (B) & 0.027 $\pm$ 0.01 & 0.041 $\pm$ 0.005 & 0.099 $\pm$ 0.02 & 0.0085 $\pm$ 0.003 & 0.028 $\pm$ 0.007 & 0.055 $\pm$ 0.02 & -0.0027 $\pm$ 0.003 & 0.012 $\pm$ 0.003 & 0.019 $\pm$ 0.01 & -0.014 $\pm$ 0.003 & 0.00098 $\pm$ 0.003 & -0.0047 $\pm$ 0.005  \\  
$RFF$ (R) & 0.075 $\pm$ 0.02 & 0.041 $\pm$ 0.005 & 0.099 $\pm$ 0.02 & 0.011 $\pm$ 0.003 & 0.028 $\pm$ 0.007 & 0.055 $\pm$ 0.02 & -0.00033 $\pm$ 0.002 & 0.012 $\pm$ 0.003 & 0.019 $\pm$ 0.01 & -0.013 $\pm$ 0.002 & 0.00098 $\pm$ 0.003 & -0.0047 $\pm$ 0.005  \\ \hline 
$\bar{\mu}_{e,r'}$ (B) & 21.3 $\pm$ 0.4 & 21.5 $\pm$ 0.3 & 21.7 $\pm$ 0.3 & 23.3 $\pm$ 0.2 & 23.0 $\pm$ 0.1 & 22.7 $\pm$ 0.4 & 24.7 $\pm$ 0.2 & 24.3 $\pm$ 0.1 & 24.1 $\pm$ 0.2 & 25.4 $\pm$ 0.2 & 24.9 $\pm$ 0.1 & 24.6 $\pm$ 0.1  \\  
$\bar{\mu}_{e,r'}$ (R) & 21.8 $\pm$ 0.3 & 21.5 $\pm$ 0.3 & 21.7 $\pm$ 0.3 & 23.4 $\pm$ 0.2 & 23.0 $\pm$ 0.1 & 22.7 $\pm$ 0.4 & 24.5 $\pm$ 0.1 & 24.3 $\pm$ 0.1 & 24.1 $\pm$ 0.2 & 25.3 $\pm$ 0.1 & 24.9 $\pm$ 0.1 & 24.6 $\pm$ 0.1  \\ \hline 
\end{tabular}
}
\end{table*}
\end{landscape}

\end{appendix}
\end{document}